\documentclass[twocolumn,showpacs]{revtex4-2}

\usepackage[dvipdfmx]{graphicx}

\usepackage{color}

\begin{document}  

\title{Persistent current and orbital magnetization along a valley-contrasting junction\\
 in bilayer graphene in a magnetic field}
\author{K. Shizuya}
\affiliation{Yukawa Institute for Theoretical Physics\\
Kyoto University,~Kyoto 606-8502,~Japan }

\begin{abstract}
In a magnetic field bilayer graphene hosts an octet of pseudo-zero-mode electron levels 
nearly degenerate in orbitals $n=(0,1)$, valleys and spins.
They split in valleys by electrostatic gating.
In gated bilayer graphene, in which the interlayer bias is set up to flip sign across a line, 
one has a line junction that traps a portion of pseudo-zero-mode electrons 
inside the insulating bulk band gap, giving rise to electron states 
localized along the junction, known as kink states. 
A close look is made into the spectra and electromagnetic response of such kink states. 
There are two species of valley current associated with them, 
a drift current driven by the bias gradient and a circulating current coming from cyclotron motion.
It turns out that they both flow in essentially the same direction, 
with the circulating current exhibiting a magnetic character distinct from those of other higher levels.  
In equilibrium they spatially circulate within the kink states, 
creating a quasi-one-dimensional channel of orbital magnetization. 
The electric control of the orbital magnetization and valley currents 
via a network of gated junctions will find useful applications in valley electronics.  

\end{abstract} 

\maketitle

\section{Introduction}

Graphene, a single layer of carbon atoms, hosts Dirac electrons as charge carriers and 
exhibits fascinating electronic properties. 
Stacking of graphene creates additional quantum effects and  further enriches its functionality, 
such as a tunable band gap in bilayer graphene~\cite{MF,NMMF,OBSH,CNMP,NCGP,ZTGH}.

The "relativistic" character of Dirac electrons becomes more apparent in a magnetic field.
In a magnetic field bilayer graphene supports, at the lowest Landau level, 
an octet $( 2_{\rm orbital}\times 2_{\rm valley}\times 2_{\rm spin})$ 
of zero-mode levels
nearly degenerate in orbitals $n=(0,1)$ as well as in valleys and spins~\cite{MF}. 
This orbital degeneracy has a topological origin~\cite{NS} in the index of the Dirac Hamiltonian.

These zero-mode levels are generally sensitive to interaction and perturbations and,
in real samples, evolve into pseudo-zero-mode (PZM) levels, 
or a variety of broken-symmetry quantum Hall states~\cite{BCNM,KS_Orbital_LS,FMY, ZCJ,WAF}.
A uniform interlayer bias applied externally acts on the two valleys $(K, K')$ 
with opposite signs and opens a tunable valley band gap,
along with a slight splitting of $n = (0,1)$ orbitals in each valley.
The pseudo-zero modes are thus naturally valley-polarized, 
with positive-energy modes in one valley 
and negative-energy ones in another valley.

Of particular interest is the case of an interlayer bias that flips sign locally
across a line on a sample~\cite{MBM}.
A bias of this type, realized by use of dual split-gate devices, 
leads to a line junction that contrasts valley-polarized PZM electrons 
in two adjacent domains. 
Such an interface traps, as the band gap closes locally, 
a portion of PZM levels within the insulating bulk band gap
and gives rise to electron (or hole) states, localized along the junction 
and accompanied by chiral valley currents counterflowing 
in different valleys~\cite{MBM,QJNM,ZPFP}.
These quasi one-dimensional (1D) states, known as kink states, 
are robust against weak disorder and the transport of injected current through them 
is nearly ballistic, as discussed in theory~\cite{QJNM} 
and observed in experiment~\cite{LWMZ,LZYZ,CZLQ,HFWT}.
The kink states are readily turned on and off by a control of gating. 
The electrical control of valley-current paths through the network 
of kink states attracts serious attention in the long-standing pursuit of 
valley electronics~\cite{RTB,XYN, GSYK,SCMS,SYBW}.

In Hall samples electrons support two species of persistent current circulating along the sample edges,
a drift current driven by the edge field and a circulating current coming from cyclotron motion,
as speculated long ago in connection with Landau diamagnetism~\cite{LP}. 
Normally they flow in opposite directions along the same edge, 
leading to oscillations of magnetization in the interior
upon varying a magnetic field, i.e., the de Haas-van~Alphen (dH-vA) effect.  
Experimentally global transport measurements accumulated evidence 
for the edge states~\cite{FKHB,WvK,LKK,SBHM,ASFA,MBA}.
Still the associated local currents, 
though actively discussed in theory~\cite{Halperin,MS,HT,Buttiker,AHK,WT,GV,IshFuk},
defied detection for years.
Only recently has it become possible, by use of  a nanoscale magnetometer, 
to directly image~\cite{UKBL} local currents in graphene, 
which indeed alternate in direction along inner edge-like configurations, 
as predicted sometime ago in a model calculation by Geller and Vignale~\cite{GV}.   
Observation of current distributions in Hall samples certainly opens a way 
to explore many-body phenomena not directly accessible by transport measurements.

In view of such developments, we examine, in the present paper, 
some observable electric and magnetic features 
of the kink states in equilibrium in the quantum Hall regime.
In our analysis, kink states are formulated  in an algebraic way via diagonalization of the PZM levels 
under a given interlayer bias, 
with their spectra and associated current distributions fixed analytically. 

In the kink states the circulating current turns out to flow in essentially the same direction as the drift one,
i.e.,  they flow in one-way on each side of the junction.  
This comes from an unusual magnetic response of the PZM electrons
while the  dH-vA oscillations are still present in other higher Landau levels. 
We further point out that the kink states, in which current circulates in equilibrium, 
are necessarily accompanied by localized orbital magnetization.
A network of valley-contrasting junctions set up on a bilayer sample will provide
a network of narrow orbital magnetization paths controllable by electric gating.

In Sec.~II we review some basic features of the Landau levels of Dirac electrons in bilayer graphene.
In  Sec.~III we determine the spectra of the kink states formed along a valley-contrasting junction. 
In  Sec.~IV, we employ a uniform interlayer bias and present, in the framework of a ($W_{\infty}$) gauge theory, 
an efficient way to derive real-space distributions of current and orbital magnetization in bilayer graphene.
In Sec.~V this analysis is generalized to the kink states of our main interest, with a careful look into
their unusual magnetic features.  
Section~VI is devoted to a summary and discussion. 
\\

\section{bilayer graphene}

The electrons in bilayer graphene are described 
by four-component spinor fields on the four inequivalent sites 
$(A,B)$ and $(A',B')$ in the bottom and top layers of graphene.
Their low-energy features are governed 
by the two Fermi points $K$ and $K'$ in the Brillouin zone. 
The intralayer coupling
$\gamma_{0} \equiv \gamma_{AB} \sim 3$\, eV
is related to the Fermi velocity 
$v_{\rm F} = (\sqrt{3}/2)\, a_{\rm L}\gamma_{0}/\hbar \sim 10^{6}$~m/s 
in monolayer graphene.
Interlayer hopping via the dimer coupling
$\gamma_{1} \equiv \gamma_{A'B} \sim 0.4$\,eV 
makes the spectra quasi-parabolic~\cite{MF} 
in the low-energy branches $|\epsilon| <\gamma_{1}$.
A key feature of bilayer graphene is that 
an externally applied interlayer bias $U_{z}$ opens a tunable band gap~\cite{MF,OBSH}.

Let us place bilayer graphene in a strong magnetic field 
$B_{z} = B>0$ normal to the sample plane, 
with the vector potential $(A_{x}, A_{y}) = (-B y, 0)$.
We set momenta $p_{i}\rightarrow \Pi_{i} = p_{i} + e A_{i}$ and rescale 
$\Pi_{x} - i\Pi_{y} = -(\sqrt{2}/\ell)\,  Z $
 and $\Pi_{x} + i\Pi_{y} = -(\sqrt{2}/\ell)\,  Z^{\dag} $
so that $[Z, Z^{\dag}]=1$, with the magnetic length $\ell = 1/\sqrt{eB}$.
Here $Z= (Y + iP)/\sqrt{2}$
in terms of $Y= (y-y_{0})/\ell$ and $P = \ell p_{y} = -i\ell \partial_{y}$, with  $[Y, P]=i$
and the center coordinate $y_{0} \equiv \ell^2 p_{x}$.

The effective Hamiltonian for bilayer graphene with leading couplings 
$(\gamma_{0}, \gamma_{1})$ and a (uniform) interlayer bias $U_{z}$ 
is written as~\cite{MF,NCGP}
\begin{eqnarray}
H &=&\! \int\! d^{2}{\bf x}\, \Big[ (\Psi^{K})^{\dag}\, {\cal H}^{K} \Psi^{K}
+ (\Psi^{K'})^{\dag}\, {\cal H}^{K'}\, \Psi^{K'}\Big], 
\nonumber\\
{\cal H}^{K} &=& \omega_{c}\left(
\begin{array}{cccc}
\mu & 0 & 0  &-Z\\
 0 & -\mu & -Z^{\dag} & 0\\
0 & -Z & - \mu & g \\
-Z^{\dag} & 0 & g  & \mu \\
\end{array}
\right),
\label{H_free}
\end{eqnarray}
where  $\Psi^{K} = (\psi_{B'}, \psi_{A}, \psi_{B}, \psi_{A'})^{\rm t}$
denotes the electron field in valley $K$, with $(A, B, \cdots)$
referring to the associated sublattices.
In ${\cal H}^{K}$,
\begin{eqnarray}
\omega_{c} &\equiv& \sqrt{2}v_{F}/\ell \approx 36.3 \times v_{\rm F}[10^{6}{\rm m/s}]\, \sqrt{B[{\rm T}]}\ {\rm meV},
\nonumber\\
g &\equiv& \gamma_{1}/\omega_{c} \  {\rm and}\ \ 
\mu  \equiv {\textstyle{1\over{2}}} U_{z}/\omega_{c}
\end{eqnarray}
represent $(\gamma_{0}, \gamma_{1}, U_{z})$ in rescaled form; 
$\omega_{c}$ is the characteristic cyclotron energy in monolayer graphene.
Here, ${\cal H}^{K}$ is diagonal in the (suppressed) electron spin.
Theoretical calculations~\cite{JM} 
suggest $v_{\rm F} \approx 0.845 \times 10^{6}\, {\rm m/s}$ and   
$\gamma_{1} \approx 361\, {\rm meV}$, which lead to 
\begin{equation}
\omega_{c}\approx 137\,{\rm meV},  g \approx 2.63 \ \  {\rm at}\  B=20{\rm T}.
\end{equation}
We employ these for numerical estimates later.

The Hamiltonian ${\cal H}^{K'}$ in another valley is given by ${\cal H}^{K}$ 
with the sign of $U_{z} = 2\omega_{c} \mu$ reversed, 
to be denoted as  ${\cal H}^{K'} = {\cal H}^{K}|_{-\mu}$, 
and acts on a spinor of the form 
$\Psi^{K'} = (\psi_{A},\psi_{B'}, \psi_{A'}, \psi_{B})^{\rm t}$, i.e., 
with the top and bottom layers interchanged $(A, B)  \leftrightarrow (B', A') $.
In what follows, we mainly refer to valley $K$, suppress valley indices,  
and specify them, when appropriate, by superscripts $(K, K')$.

The eigenmodes of ${\cal H}^{K}$ are labeled by integers $N\equiv |n| =(0,1,2,\cdots)$,
to be called $\lq\lq$sectors" below,  
and $y_{0}=\ell^2p_{x}$, and have the structure 
$\Phi^{n}_{y_{0}}({\bf x}) = \langle {\bf x}| \Phi^{n}_{y_{0}}\rangle$, 
with 
\begin{equation}
| \Phi^{n}_{y_{0}}\rangle = \Big(|N-2\rangle_{y_{0}}\, b'_{n} ,|N\rangle_{y_{0}}\, c_{n},
|N -1\rangle_{y_{0}}\, b_{n}, |N-1\rangle_{y_{0}}\, c'_{n} \Big)^{\rm t}. 
\label{Phi_n_yz}
\end{equation} 
Here 
$\langle {\bf x}|N\rangle_{y_{0}} =  \langle x|y_{0}\rangle  \langle y|N\rangle$
are wave functions of conventional Hall electrons (of quadratic dispersion), 
consisting of plane waves 
$\langle x|y_{0}\rangle = e^{ix\,y_{0}/\ell^2}/\sqrt{2\pi \ell^2}$
of momentum $p_{x}$
and the harmonic-oscillator wave functions $\langle y|N\rangle =\phi_{N}(y-y_{0}\rangle$, 
with
\begin{equation}
\phi_{N}(y)
= e^{-{1\over{2}} (y/\ell)^2}\, H_{N}(y/\ell)/\sqrt{N!\, 2^{N}\sqrt{\pi}\,  \ell};
\end{equation}
$|N\rangle_{y_{0}}  =0$ for $N<0$.
(We use capital letters  for absolute values,  $n= \pm N$.) 
The coefficients ${\bf b}_{n}=(b'_{n}, c_{n}, b_{n}, c'_{n})^{\rm t}$
for each $n=\pm N$ are given by the orthonormal eigenvectors 
of the reduced matrix ${\cal H}_{N}^{\rm red}$ obtained from ${\cal H}^{K}$ 
by replacing $(Z, Z^{\dag})\rightarrow \sqrt{N-1}$ for $(b'_{n}, c'_{n})$,
and $(Z, Z^{\dag}) \rightarrow \sqrt{N}$ for $(b_{n}, c_{n})$.

For each $N\ge 2$, ${\cal H}_{N}^{\rm red}$ has rank 4 and 
we denote the four branches of  Landau-level spectra as 
$\epsilon_{-n'} < \epsilon_{-n}<0<\epsilon_{n} < \epsilon_{n'}$ 
(with $|\epsilon_{\pm n'}| \gtrsim \gamma_{1}$)
so that the index $(n, n') \in N$ reflects the sign of $\epsilon_{n}$. 
Let us denote $\epsilon_{n}$ in units of $\omega_{c}$, 
\begin{equation}
\epsilon_{n} = \omega_{c}\, e_{n}\ {\rm and}\ \epsilon_{n'} = \omega_{c}\, e_{n'}.
\end{equation}
The (dimensionless) spectra of the lower branches $n=\pm N$ are written, to $O(\mu)$, as 
\begin{eqnarray}
e_{n}\!
&=&\!  s_{n} {1\over{\sqrt{2}}}\sqrt{ g^2 + a_{N} - \sqrt{D_{N}}} 
 +{ \mu \over{\sqrt{D_{N}}}}  \sim O(1/g),
\label{En_zeroth}
\end{eqnarray} 
while $e_{n'} \sim O(g)$ of the higher branches take the same form 
with $\sqrt{D_{N}} \rightarrow - \sqrt{D_{N}}$.
Here $a_{N}= 2\, N-1$ and $D_{N}= (g^2 +a_{N})^2 -4N(N-1)$;
$s_{n} \equiv {\rm sgn}(n) = \pm 1$.
For each $e_{n}$ and $e_{n'}$ the eigenvector  ${\bf b}_{n}$ is written as
\begin{equation}
{\bf b}_{n} = c_{n} \Big(  -{1\over{g}} {\sqrt{N(N\!-\!1)}\over{e_{n}-\mu}}\,\lambda_{n}, 
1,  - {e_{n}+\mu \over{\sqrt{N}}}, { \sqrt{N}\over{g}} \lambda_{n}\Big),
\label{ev_b_n}
\end{equation}
with $\lambda_{n} \equiv 1- ( e_{n}+\mu)^2/N$; $c_{n}$ is chosen to normalize ${\bf b}_{n}$.

Of our particular concern are the $N=0$ and $N=1$ sectors.
For $n=0$,  ${\cal H}_{N=0}^{\rm red}$ has an obvious eigenmode with
\begin{equation}
\epsilon_{0}= - U_{z}/2 = -\omega_{c}\mu, \ \ {\bf b}_{0} = (0,1,0,0)^{\rm t}.
\end{equation}
For $N=1$,   ${\cal H}_{N}^{\rm red}$ has three solutions
$(\epsilon_{-1'}, \epsilon_{1},\epsilon_{1'})$.
The $n=1$ mode has the spectrum and eigenvector, to $O(\mu)$, 
\begin{eqnarray}
e_{1} &=& -\mu +{2 \mu\over{g^2 +1}} + O(\mu^3/g^6) 
\approx - {g^2 -1\over{g^2 +1}}\, \mu  ,
\nonumber\\
{\bf b}_{1}&=& 
c_{1} \Big(0,1, - 2 \mu/(g^2 +1) , 1/g \Big),\ \ \ 
\nonumber\\
c_{1} &=& g/\sqrt{g^2 +1} + O(\mu^2/g^4), 
\end{eqnarray}
while those of the $n= \pm 1'$ modes read 
\begin{eqnarray}
e_{\pm 1'} &=& \pm \sqrt{g^2+1} -\mu/(g^{2}+1),
\nonumber\\
{\bf b}_{\pm1'} 
&=&(1/\sqrt{2}) (0,  -c_{1}/g, \pm 1,  c_{1})  + O(\mu);
\end{eqnarray}
numerically, $c_{1} \approx 0.935 \sim O(1)$ for $g=2.63$.

The interlayer bias $U_{z} = 2\omega_{c} \mu$ thus shifts 
the orbital modes $n= (0 , 1)$ from zero energy, 
with $e_{0}^{K}  = -\mu$ and $e_{1}^{K} = -\mu + O(\mu/g^2)$. 
It breaks valley symmetry 
and shifts those PZM levels oppositely ($\epsilon_{n}^{K'} = -\epsilon_{n}^{K}$) in the two valleys $(K,K')$, 
opening a band gap $\sim U_{z}$. 
Their spectra are nearly degenerate in each valley, and are ordered as 
$\epsilon_{0}^{K} \lesssim \epsilon_{1}^{K} < 0 
< \epsilon_{1}^{K'} \lesssim \epsilon_{0}^{K'}$ for $\mu >0$.
In the present paper we handle both signs of $\mu$ 
and will specify the signs of $(\epsilon_{0}, \epsilon_{1})$ 
by the level index $n$ with a subscript $\pm$ attached.
Note that the $n=(0_{+}^{K'}, 1_{+}^{K'})$ levels present for $\mu > 0$ 
are physically distinct from the $(0_{+}^{K}, 1_{+}^{K})$ levels realized for $\mu<0$; 
they look similar but their wave functions reside in opposite layers, 
$(A, B)  \leftrightarrow (B', A')$. 
We suppose that $|e_{0}| =  \mu \ll e_{2}$  ($\approx 0.45$  for $g\approx 2.63$),
so that these PZM levels are well-isolated from others.

Passing to the $| \Phi^{n}_{y_{0}}\rangle$ basis 
via the expansion 
$\Psi^{K}({\bf x}) = \sum_{n, y_{0}} \langle {\bf x}| \Phi^{n}_{y_{0}}\rangle\, \psi_{n}(y_{0})$  
in $H$ yields a one-body Hamiltonian in valley $K$ of the form~\cite{ks_edgeC_W} 
\begin{eqnarray}
H &=& \int dy_{0}\sum_{m, n}
\psi_{m}^{\dag}(y_{0})\, {\cal H}^{mn}\, \psi_{n}(y_{0}),
\label{H_BG}\\
{\cal H}&=&  \omega_{c} \big\{ - b' Z c'  - c' Z^{\dag} b' - b Z c - cZ^{\dag}b
\nonumber\\ 
&&+ g (bc' + c'b) + \mu( b' b' +c' c') - \mu (bb+ cc) \big\}.\ \ 
\label{H_mn}
\end{eqnarray}
Here ${\cal H}$ represents a matrix ${\cal H}^{mn}$ 
in orbital labels $(m,n)$ which now run over all integers $n = (0_{-}, \pm1,\pm2, \cdots)$
(and over $n'$ as well); 
in what follows we adopt such matrix notation and 
frequently suppress summation over repeated labels. 
In ${\cal H}$ we have introduced condensed notation: 
For ${\cal H}^{mn}$ we interpret, e.g., 
\begin{eqnarray}
b'  Z\, c' \! &\rightarrow&    b'_{m}\,Z^{M-2,N-1} c'_{n}, 
b\, Z\, c  \! \rightarrow    b_{m}\, Z^{M-1,N} c_{n},
\nonumber\\
c\, Z^{\dag} b \! &\rightarrow&    c_{m}\, [Z^{\dag}]^{M,N-1} b_{n}, 
\nonumber\\
bc' &\rightarrow& b_{m}1^{M-1, N-1}c'_{n}, b' b' \rightarrow b'_{m}1^{M-2, N-2}b'_{n},
\label{condensed_nt}
\end{eqnarray}
as seen from $| \Phi^{n}_{y_{0}}\rangle$ in Eq.~(\ref{Phi_n_yz}); 
$M=|m|$, $N= |n|$ and  $1^{MN} \equiv \delta^{MN}$.
Noting $(b Z c)^{mn} =  \delta^{MN} \sqrt{N}\, b_{m}c_{n}$, 
$(b' Z c')^{mn} = \delta^{MN} \sqrt{N-1}\, b_{m}c_{n}$, etc.,
one readily sees that ${\cal H}$ is reduced to the diagonal spectra 
${\cal H}^{mn} = \epsilon_{n}\,  \delta^{mn}$.

\section{Valley-contrasting interface }

%%%%%%%%%%%%%%% Figure1 %%%%%%%%%
 \begin{figure}[tpb]
\begin{center}
\includegraphics[scale=.95]{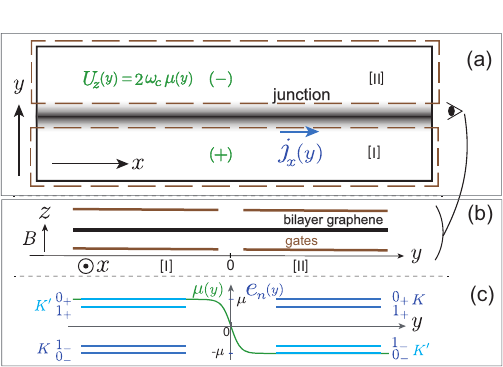}
\end{center}
\caption{ 
(a)~Schematic of a bilayer sample divided into two domains  
by electrostatic gating across a junction (shaded area). 
(b)~Cross-section view of the sample.  
(c)  Interlayer bias $\mu(y) = - \mu\, {\rm tanh} ( y /w_{\rm jc})$ with a reversal of polarity across $y=0$. 
Shown in blue are the associated PZM spectra
of valleys $K$ and $K'$ in the bulk domains $|y| \gg w_{\rm jc}$.}
\end{figure}
%%%%%%%%%%%%%%%%%%%%%%%%

Let us now introduce an interface that contrasts valley-polarized PZM electrons~\cite{MBM,LWMZ}.  
Figure~1 illustrates a Hall-bar sample we consider. 
It is homogeneous in the length direction $\{x\}$ while, in the width direction $\{y\}$, 
it is divided into two domains by electrostatic gating. 
When an almost uniform bias $U_{z}$  is chosen to flip sign
across a line $y=0$,
$U_{z} \rightarrow U_{z}(y) = 2 \omega_{c}\,  \mu (y)$,  as in Fig.~1(c), 
the band gap closes there and there emerge, 
within the insulating bulk band gap $2 \omega_{c} |e_{1}| \sim U_{z}$,
electron (or hole) states localized across and along the junction.
In this section we examine the detailed structure 
of such boundary states~\cite{MBM}, known as kink states, 
by adopting a bias of the form 
\begin{equation}
\mu (y) = - \mu\, {\rm tanh} ( y /w_{\rm jc})\ \ ({\rm with}\ \ \mu>0),
\label{PotWall}
\end{equation}
which rapidly changes from $\mu >0$ to $-\mu$ over an interval $|y| \lesssim 2 w_{\rm jc}$.

Let us start with ${\cal H}$ in Eq.~(\ref{H_mn}). Upon setting $\mu \rightarrow \mu(y)$ there, 
the  $(\mu b'b' + \cdots)$ term is replaced with
\begin{equation}
{\cal V}_{\mu} = \omega_{c} \big\{b' \mu (y) b' + c' \mu (y) c' - b\,  \mu (y) b - c\, \mu (y) c \big\},
\end{equation}  
where $b' \mu (y)\,  b' = b'_{m}[\mu (y)]^{M-2, N-2} b'_{n}$, etc. 
$({\cal V}_{\mu})^{mn} $ is no longer diagonal around $|y_{0}| \lesssim 2 w_{\rm jc}$.
In the oscillator basis $\langle y|N\rangle  = \phi_{N}(y-y_{0})$, 
the coordinate $y = y_{0} + \ell Y$ becomes a matrix
$y^{MN} = y_{0}\, \delta^{MN} + \ell\,  Y^{MN}$,
with $Y = (Z + Z^{\dag})/\sqrt{2}$ and  $Z^{MN} \equiv \langle M|Z|N\rangle= \sqrt{N}\delta^{M, N-1}$.
The bias $\mu(y)$  then turns into $\mu(y_{0} + \ell Y)^{MN}$, 
a function local in center coordinate $y_{0}$ and a matrix  in "sector" labels $(M,N) \in (0, 1, 2, \dots)$.
Its matrix portion is isolated as
\begin{equation}
\mu (y) = e^{\ell Y \partial_{y_{0}}} \mu(y_{0}) \equiv e^{\ell Y ip} \mu (y_{0}),
\end{equation}
where $ip= \partial_{y_{0}}$ stands for a derivative acting on the local function  $\mu (y_{0})$.
(For clarity, we write $[{\cal O}]^{MN}$ for ${\cal O}^{MN} =\langle M|{\cal O}|N\rangle$ below.) 
One can rewrite 
\begin{equation}
 e^{Y i\ell p}= \gamma_{p}\,  e^{Z^{\dag} i \ell p/\sqrt{2}} e^{Z i\ell p/\sqrt{2}} 
 \equiv  \gamma_{p} f_{p},
 \end{equation}
where $\gamma_{p} = e^{-{1\over{4}} \ell^2 p^2}$  
and $f_{p}^{MN} =\langle M| f_{p} |N \rangle$ are expressed 
in terms of the associated Laguerre polynomials,
\begin{equation}
f_{p}^{MN} = \sqrt{N!/M!}\,  (i\ell p/\sqrt{2})^{M-N} L_{N}^{M-N}(\ell^2 p^2/2);
\label{f_p_MN}
\end{equation}
$f_{p}^{NM} = (f_{-p}^{MN})^{*}$; $f^{00}_{p}=1$, $f^{10}_{p} = f^{01}_{p} = i\ell p/\sqrt{2}$,
$f^{11}_{p} = 1- {1\over{2}}\ell^2 p^2$, etc. 
In this way, $\mu(y)$ is naturally expanded in a normal-ordered series of $(Z^{\dag}, Z)$ 
and in multipoles of $\mu (y_{0})$, and one can simply write 
\begin{eqnarray}
[\mu (y)]^{MN}&=&  \int_{-\infty}^{\infty}\!\!dy\, \mu (y)\,   \phi_{M}(y-y_{0}) \phi_{N}(y-y_{0}).
\nonumber\\
&=& f^{MN}_{p} \gamma_{p}\,  \mu (y_{0}) \equiv f^{MN}_{p} m_{0}(y_{0})
 \label{eipy_phiM}
\end{eqnarray}
as multipoles of the lowest one 
$m_{0}(y_{0})\equiv [\mu(y)]^{00}$, i.e.,  
\begin{equation}
m_{0}(y_{0}) \equiv \gamma_{p} \mu(y_{0}) = \int\! dy\,  \mu(y)\,   |\phi_{0}(y-y_{0})|^2.
 \label{mu_y_to_yzero}
 \end{equation}
The  exponential (derivative) factor $\gamma_{p} = e^{-{1\over{4}} \ell^2 p^2}$ 
works to suppress short-wavelength variations  ($|p| \gg \ell$) in $\mu(y)$, 
and $m_{0}(y_{0})$ varies smoothly on the scale of  $O(\ell)$.
Even a step function 
$\mu (y)|_{w_{\rm jc} \rightarrow 0}  =  - \mu\,  {\rm sgn}(y)$
yields an expression  
$m_{0}(y_{0}) = -\mu\,  {\rm erf}(y_{0}/\ell)$  smooth in $y_{0}$.
We  hereafter treat $m_{0}(y_{0})$ as a basic $\lq\lq$order parameter".

Replacing $\mu \rightarrow \mu (y)$ fundamentally modifies the $O(\mu)$ spectra 
of the PZM levels $n=(0, 1)$.
To fix them  one first has to rediagonalize 
${\cal H}^{K}$ of Eq.~(\ref{H_free}) [with $\mu\rightarrow \mu(y)$] in each $N$ sector. 
In the $N=0$ sector, $e_{0}= - \mu$ is promoted to $e_{0}(y_{0}) = -[\mu (y)]^{00} = -m_{0}(y_{0})$.
In the $N=1$ sector,  the $n=1_{-}$ mode acquires, to $O(\mu)$,
the spectrum 
\begin{equation}
e_{1}(y_{0}) = -\{g^2 [\mu(y)]^{11} - [\mu(y)]^{00}\}/(g^2 +1)
\label{e_one_y}
\end{equation}
and the eigenvector ${\bf b}_{1}$
with  $c_{1} = g/\sqrt{g^2+1}$ and $c'_{1} =c_{1}/g$, as before, 
while 
\begin{equation}
b_{1} = -c_{1} \{ [\mu(y)]^{11} + [\mu(y)]^{00}\} /(g^2 +1)
\label{b_one_y}
\end{equation}
now depends on $y_{0}$.
In addition, the $N=\{0,1\}$ sectors are coupled through $[\mu (y)]^{10} = f^{10}_{p} m_{0}(y_{0})$.

We expand $[\mu (y)]^{MN}$ in multipoles of $m_{0}(y_{0})$ up to $O(\partial_{y_{0}}^2 m_{0})$,
and find that the PZM sector $\ni n=(0,1)$ (in valley $K$) is governed by the Hamiltonian 
\begin{equation}
H= \int dy_{0} \psi_{m}^{\dag}(y_{0})\, {\cal H}_{\rm PZ}^{mn}\, \psi_{n}(y_{0}),
\label{H_PZM}
\end{equation}
with ${\cal H}_{\rm PZ}^{mn} \equiv \omega_{c} h^{mn}$ written as
\begin{eqnarray}
h^{00} &=& - m_{0}(y_{0}) =  -Q(y_{0}) + { c_{1}\over{2\sqrt{2}}}\,  d R(y_{0}),
\nonumber\\
h^{11} 
&=& - {g^2\! -1\over{g^2+1}}\, m_{0}(y_{0}) -{1\over{2}} (c_{1})^2d^2 m_{0}(y_{0})
\nonumber\\
&=& - \{1- 2 (c_{1}/g)^2 \}\, Q(y_{0}) + {c_{1} \over{2\sqrt{2}}}\,  d R(y_{0}),
\nonumber\\
h^{10}&=& h^{01} = - {c_{1}\over{\sqrt{2}}}\,  d m_{0}(y_{0}) = (c_{1}/g)^2 R(y_{0}),
\label{h_pzm} 
\end{eqnarray}
where $c_{1} = g/\sqrt{g^2+1}$ and $d\equiv i\ell p = \ell \partial_{y_{0}}$.
Here, for later convenience, the following notation is introduced, 
\begin{eqnarray}
Q(y_{0}) &=& m_{0}(y_{0}) - {\textstyle {1\over{4}}}\, g^2 d^2m_{0}(y_{0}),  
\nonumber\\
R(y_{0}) &=& - (g^2/ \sqrt{2} c_{1})  d m_{0}(y_{0}).
\label{QR_rep}
\end{eqnarray}
From now on we denote $d = \ell \partial_{y_{0}}$, and
suppress the magnetic length 
$\ell \rightarrow 1$ and recover it, when appropriate.

The diagonalization of $h^{mn}$ is viewed as a spinor rotation of the  $n= (0, 1)$ levels,
\begin{eqnarray}
\psi &=& (\psi_{1}, \psi_{0})^{\rm t} \rightarrow \hat{\psi} 
= (\hat{\psi}_{1}, \hat{\psi}_{0})^{\rm t} = U \psi, 
\nonumber\\
U&=&e^{ i\theta (y_{0})  \sigma_{2} /2} =  \cos (\theta/2)  + i\sigma_{2} \sin (\theta/2), 
\label{Rotation_U}
\end{eqnarray}
with a Pauli matrix $\sigma_{2}$.
The angle $\theta(y_{0})$ is fixed as
\begin{eqnarray}
&&\{\cos \theta(y_{0}),\sin \theta (y_{0})\} = \{Q(y_{0}), R(y_{0})\} /\sqrt{D(y_{0})}, 
\nonumber\\
&&D(y_{0}) = Q(y_{0})^{2} + R(y_{0})^{2},
\label{MixingAngle}
\end{eqnarray}
where we have set $m_{0} (y_{0}) \rightarrow \mu > 0$ 
and $\theta (y_{0}) \rightarrow 0$ for $y_{0} \ll 0$.
This leads to the diagonal Hamiltonian 
\begin{equation}
H = \int dy_{0} \hat{\psi}_{n}^{\dag}(y_{0})\, {\cal E}_{n}(y_{0})\, \hat{\psi}_{n}(y_{0})  
\end{equation}
with the spectra  ${\cal E}_{n}(y_{0})/\omega_{c} \equiv \hat{e}_{n}(y_{0}) = (Uh U^{\dag})^{nn}$ 
and
 \begin{eqnarray}
\hat{e}_{1}(y_{0})  
&=& e_{\rm mid}(y_{0}) + (c_{1}/g)^2 \sqrt{D(y_{0})} ,
\nonumber\\
\hat{e}_{0}(y_{0}) &=& e_{\rm mid}(y_{0}) - (c_{1}/g)^2 \sqrt{D(y_{0})}.
\nonumber\\
e_{\rm mid}(y_{0}) &=&- (c_{1})^2Q(y_{0}) + {c_{1}\over{2\sqrt{2}}}\, dR(y_{0}) ,
\label{Kink_Spectra}
\end{eqnarray}

%%%%%%%%%%%%%%% Figure2 %%%%%%%%%
 \begin{figure}[tpb]
\begin{center}
\includegraphics[scale=.85]{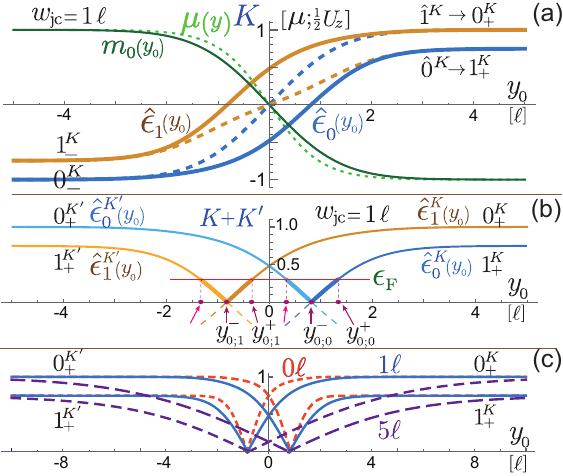}
\end{center}
\caption{
 (a)  PZM spectra in valley $K$ under a sharp bias $\mu(y)$ with $w_{\rm jc} = 1 \ell$.  
 Levels $(1_{-}^{K}, 0_{-}^{K})$ in the lower domain $y_{0} \ll 0$ get mixed across the junction at $y_{0} \sim 0$
 and turn into $(0_{+}^{K}, 1_{+}^{K})$ for $y_{0} \gg 0$, avoiding thereby level crossing.
 Another valley $K'$ also supports the same spectra with inverted signs, 
with  orbital change  $(0_{+}^{K'}, 1_{+}^{K'}) \rightarrow (1_{-}^{K'}, 0_{-}^{K'})$.
(b)~Electron spectra above the charge-neutral ground state in full valleys $K+K'$.
Four channels (per spin) of active orbitally-mixed electron modes are trapped within the bulk band gap. 
(c)~The level spectra scarcely change as the kink gets steeper $w_{\rm jc} =1\ell \rightarrow 0$ while they 
spread in width with increasing $w_{\rm jc}$. 
}
\end{figure}
%%%%%%%%%%%%%%%%%%%%%%%%

Figure 2(a)  shows the spectra $\{ \hat{e}_{0}(y_{0}), \hat{e}_{1}(y_{0}) \}$ in valley $K$ 
under a {\it sharp} kink bias $\mu(y) = -\mu\, \tanh (y_{0}/w_{\rm jc})$, with $w_{\rm jc} =1 \ell$ 
and for the choice of  $g =2.63$. 
There the $(1_{-}, 0_{-})$ levels, starting with flat spectra  $(e_{1_{-}}, e_{0_{-}})$
[and $\theta (y_{0}) =0$] on the lower side $y_{0}\ll 0$, 
get mixed across the junction [with $\theta (0) \sim \pi/2$] 
and turn into the $(0_{+} ,1_{+})$ levels with spectra  $(e_{0_{+}}, e_{1_{+}})$ 
for $y_{0}\gg 0$ [where $\theta (y_{0})\rightarrow \pi$ and  $m_{0}(y_{0}) \rightarrow -\mu$].
The PZM levels thus undergo, via mixing, an orbital change 
$\{ \psi_{1},  \psi_{0}\} \rightarrow \{\psi_{0}, -\psi_{1}\}$ or 
$(1_{-}^{K}, 0_{-}^{K}) \rightarrow (0_{+}^{K}, 1_{+}^{K})$ across the junction,
avoiding level crossing with a finite gap.  
Clearly this orbital mixing is characterized by a non-Abelian Berry phase~\cite{WZ}
\begin{equation}
-iU^{\dag}dU= d\theta (y_{0}) \sigma_{2}/2,
\end{equation}
which changes by $\Delta \theta = \pi$. 
Note that, if orbital coupling $h^{01}$ were ignored, 
the two levels would simply intersect at the junction; see dashed curves in Fig.~2(a).

There is another set of PZM levels, which starts with $n=(0_{+}, 1_{+})$ in valley $K'$ for $y_{0} \ll 0$.
Their spectra simply differ in sign, 
$\hat{e}^{K'}_{n} (y_{0}) = - \hat{\epsilon}^{K}_{n} (y_{0})$, 
by setting $\mu \rightarrow -\mu$ (i.e., $m_{0}\! \rightarrow\! - m_{0}$ and  $\sqrt{D}\! \rightarrow\! -\sqrt{D}$)
in Eqs.~(\ref{h_pzm}) - (\ref{Kink_Spectra}).

In the $\lq\lq$relativistic" electron-hole picture of bilayer graphene 
the ground state of electrons is a charge-neutral surface of energy $\epsilon =0$, 
or the Dirac sea with all negative-energy levels filled in both valleys. 
In this picture,  among the PZM  electrons,
only those of levels $(0_{+}^{K'}, 1_{+}^{K'})$ and $(0_{+}^{K}, 1_{+}^{K})$
are seen above the Dirac sea in the bulk regions, as depicted in Fig.~2(b), 
and they $\lq\lq$appear" to undergo an orbital and valley change 
$(0_{+}^{K'}, 1_{+}^{K'}) \rightarrow (1_{+}^{K}, 0_{+}^{K})$ across the junction.
The junction thus contrasts valley-polarized PZM electrons in adjacent domains.
There the kink of bias $\mu(y)$ provides local potentials that confine four channels 
$( \hat{0}_{+}^{K'}, \hat{1}_{+}^{K'}, \hat{0}_{+}^{K}, \hat{1}_{+}^{K})$ (per spin) 
of active electron modes 
around the zeros of $\hat{\epsilon}_{0}(y_{0})$ and $\hat{\epsilon}_{1}(y_{0})$,
i.e., $y_{0;0}^{-}$ and $y_{0;1}^{-}$.  
[Here $n=(\hat{0}, \hat{1})$ refer to the mixed levels.] 
They form, when the Fermi energy $\epsilon_{\rm F}$ is set to cross them, 
electron kink states localized along the junction.

As seen from Fig.~2(c), the kink states spread in width 
with increasing $\epsilon_{\rm F}$ and width $w_{\rm jc}$.  
The $\epsilon=0$ positions $\{y_{0;0}^{-},  y_{0;1}^{-}\}$ barely move 
while the level gap $\hat{e}_{1}(0) - \hat{e}_{0}(0)$ at the junction 
decreases with increasing $w_{\rm jc}$.
For $w_{\rm jc} \gtrsim 5\ell$, $m_{0}(y_{0})=\gamma_{p}\mu(y_{0})$ 
scarcely deviates from $\mu (y_{0})$, and one finds that
\begin{eqnarray}
&&y_{0;0}^{-} = - y_{0;1}^{-} \approx {1\over{\sqrt{2} c_{1}}} \ell \approx 0.76 \ell \ \ {\rm for}\   g = 2.63, 
\nonumber\\
&&\hat{e}_{1}(0) - \hat{e}_{0}(0) \approx \sqrt{2}c_{1}  |dm_{0}(0)| \approx \sqrt{2} c_{1}\mu\,  \ell / w_{\rm jc}.
\label{Focuses}
\end{eqnarray}

In Sec.~V we further examine observable features of  the kink states. 
As a preliminary, the next section presents a systematic way to handle 
the electromagnetic response of Hall electrons.

\section{Electromagnetic response of Dirac electrons}

In this section we study the electromagnetic response of Dirac electrons in bilayer graphene.
We write a current probe ${\bf a} = (a_{x}, a_{y})$
in a dimensionless form $v_{i} =e \ell\, a_{i}$,
set $v({\bf x})= \{v_{y}({\bf x}) + i v_{x}({\bf x})\}/\sqrt{2}$  and 
replace, in the original Hamiltonian ${\cal H}^{K}$  in Eq.~(\ref{H_free}), 
$Z \rightarrow Z + i v({\bf x})$ and $Z^{\dag} \rightarrow Z^{\dag} - iv^{\dag}({\bf x})$.  
With a static potential $-e A_{0}({\bf x})$ added, 
\begin{equation}
{\cal H}^{K}|_{Z\rightarrow Z+iv, Z^{\dag} \rightarrow Z^{\dag}-iv^{\dag}} - eA_{0}
\label{H_vA}
\end{equation}
is now regarded as ${\cal H}^{K} =  {\cal H}^{K}[v, A_{0}; \mu]$.
As we consider a sample homogeneous  in  the $x$ direction
it suffices to adopt potentials $\{a_{x}(y), A_{0}(y) \}$ (with $a_{y}\rightarrow 0$) 
that vary only in the width direction $\{y\}$. 
 One can then detect an $x$-averaged current $j_{x}(y) = (1/L_{x}) \int dx\, j_{x}(x,y)$
(with $L_{x} = \int dx$) driven by local potentials $\{\mu(y), A_{0}(y) \}$
and a vertical magnetic field  
$b_{z}^{({\bf a)}} \equiv \nabla \times {\bf a}  \rightarrow -\partial_{y}a_{x}(y)$.

Let us now look at this $4\times 4$ Hamiltonian 
${\cal H}^{K}[v, A_{0}; \mu]$ carefully. 
Note first that ${\cal H}^{K}$ is unitarily equivalent to 
$-{\cal H}^{K}$ with the signs of $\mu$ and $A_{0}$ reversed~\cite{KS_Orbital_LS}, 
\begin{equation}
C^{\dag}{\cal H}^{K}\, C = - {\cal H}^{K}|_{- \mu, -A_{0}},
\end{equation}
with $C= {\rm diag}(-1,1,-1,1)$.
The electron and hole bands are therefore intimately related within a valley.
One can also pass to another valley $K'$ 
by setting $\mu \rightarrow -\mu$, i.e., 
${\cal H}^{K'} = {\cal H}^{K}|_{-\mu}$.
The unitary equivalence
\begin{equation}
{\cal H}^{K'} =  {\cal H}^{K}|_{-\mu}
 \stackrel{C}{\sim} - {\cal H}^{K}|_{-A_{0}}
 \label{unitary_equiv}
 \end{equation}
implies, in view of the discreteness of the (exact) level spectra 
${\cal E}_{n} = {\cal E}_{n}(y_{0})$ for $\mu (y) \not=0$, 
the following relation, 
\begin{eqnarray}
{\cal E}^{K'}_{n} &\stackrel{ n \not= (0,1)}{=}& {\cal E}^{K}_{n}|_{-\mu} 
= -{\cal E}^{K}_{-n}|_{-A_{0}}, 
\nonumber\\
{\cal E}^{K'}_{n_{+}} &\stackrel{ n = (0,1)}{=}&  {\cal E}^{K}_{n_{-}}|_{-\mu}  
=  -{\cal E}^{K}_{n_{-}}|_{-A_{0}}.  
\label{ConjugatedSpec}
\end{eqnarray}
Thus the spectra of valley $K'$ are immediately derived from $\{{\cal E}^{K}_{n}(y_{0})\}$ of valley $K$.

Let us pass from $\Psi^{K}({\bf x})$ to $\psi_{n}(y_{0})$, as in Eq.~(\ref{H_BG}), 
and consider the Hamiltonian
$H = \int dy_{0}\,  \psi_{m}^{\dag}(y_{0})\, {\cal H}^{mn}\, \psi_{n}(y_{0})$, 
now with ${\cal H} ={\cal H}[v,A_{0};\mu]$.
The electric current $j_{x}(y) = - (1/L_{x}) \delta H/\delta a_{x}(y)$ is read from the Hamiltonian $H$
through terms linear in potential $a_{x}$.
The external fields $(a_{x}, A_{0}, \mu)$, upon introduction, induce level mixing in ${\cal H}^{mn}$. 
To derive a current $j_{x}$ in each diagonal level $\{n\}$ 
one thus has to rediagonalize ${\cal H}^{mn}$ in orbitals $(m,n)$,  at least to $O(a_{x})$,
by a suitable unitary transformation,
$\psi \rightarrow \hat{\psi} = U\psi$, 
so that $H = \int dy_{0} \hat{\psi}_{n}^{\dag}(y_{0})\, {\cal E}_{n}\, \hat{\psi}_{n}(y_{0})$
with $(U{\cal H}U^{\dag})^{mn} =\delta^{mn} {\cal E}_{n}$. 
Then Eq.~(\ref{ConjugatedSpec}) tells us that currents in valleys $K$ and $K'$  
are intimately related, 
\begin{eqnarray}
j_{n}^{K'}(y) &=&  j_{n}^{K}(y)|_{-\mu} = - j_{-n}^{K}(y)|_{-A_{0}},
\nonumber\\
j_{n_{+}}^{K'}(y) &=&  j_{n_{-}}^{K}(y)|_{-\mu} =- j_{n_{-}}^{K}(y)|_{-A_{0}}, 
\label{jK_vs_jKp}
\end{eqnarray} 
in obvious notation,
where $j_{n}^{K}(y) \propto  - \{\delta/\delta a_{x}(y)\}  {\cal E}^{K}_{n}(y_{0})$
stands for the current $j_{x}(y)$ carried by the $(n,y_{0})$ mode in valley $K$, etc.

There are in general two types of current in a magnetic field, 
a drift current, denoted as $j^{\rm (d)}$, driven by local fields $\{\mu (y), A_{0}(y)\}$ 
and a circulating current, denoted as  $j^{\rm (c)}$, coming from cyclotron motion.  
(i) Let us first extract from Eq.~(\ref{jK_vs_jKp}) 
 the drift component $j^{\rm (d)}_{n} \propto O(A_{0})$  driven by $A_{0}$, i.e., the ordinary Hall current,
\begin{equation}
j_{n}^{{\rm (d)};K'} =  j_{n}^{{\rm (d)};K}|_{-\mu} =  j_{-n}^{{\rm (d)};K}
\  \ {\rm for}\ \   j^{{\rm (d)}}_{n} \propto O(A_{0}).
\label{jhall_KKp}
\end{equation}
The Hall current $j^{{\rm (d)}}_{n} \propto O(A_{0})$ therefore
flows in the same direction for all $n$ in both valleys. 
(ii)~In contrast, another component $j^{\rm (d)}_{n} \propto O(\mu)$ 
driven by the bias $\mu(y)$, 
\begin{equation}
j^{{\rm (d)};K'}_{n}  = - j^{{\rm (d)};K}_{n},\ 
j^{{\rm (d)};K'}_{n_{+}}  = - j^{{\rm (d)};K}_{n_{-}}  \propto O(\mu),
\label{jd_K_Kp}
\end{equation}
flows in opposite directions in different valleys. 
The valley-contrasting junction thus realizes the valley Hall effect~\cite{XYN} 
with counterflowing valley currents.

(iii)~The circulating currents $j^{\rm (c)}_{n}$, detected by a magnetic probe $\propto \partial_{y}a_{x}$,
are present in the sample interior. 
Let us pick them up by setting $A_{0} \rightarrow  0$, 
\begin{equation}
j^{{\rm (c)};K'}_{n}= j^{{\rm (c)};K}_{n}|_{-\mu}=   - j^{{\rm (c)};K}_{-n},\ \
 j^{{\rm (c)};K'}_{n_{+}} =  - j^{{\rm (c)};K}_{n_{-}}. 
 \label{eh-conj_jc}
\end{equation}
It is seen that positive- and negative-energy Dirac electrons of level $\pm n$ carry current of opposite signs. 
This implies, in a semiclassical picture, that they mutually orbit in opposite directions in a magnetic field.

In practice, it is sufficient to diagonalize $H$ to $O(a_{x})$ 
to construct the current $j_{x}(y) = j^{\rm (d)}(y) +j^{\rm (c)}(y)$.
There is an efficient way to achieve such diagonalization to $O(a_{x})$
by use of a gauge transformation~\cite{ks_edgeC_W}.
In this section, we present it for a {\em uniform} bias $\mu >0$ 
and examine the effect of the static potential $A_{0}(y)$ in bilayer graphene.

To derive the current $j_{x}(y)$ in real $y$ space, 
it is necessary to keep the full multipoles of the probe $v_{x}(y) = e\ell a_{x}(y)$.
We thus adopt 
\begin{equation}
u_{x}(y_{0}) \equiv \gamma_{p} v_{x}(y_{0}) = \int\! dy\, v_{x} (y)  |\phi_{0}(y-y_{0})|^2
\label{vx_real-vx}
\end{equation}
as our basic field and set $[v_{x}(y)]^{MN} =f^{MN}_{p} u_{x}(y_{0})$.
Let us now consider a unitary transformation $G  =  G^{MN}$ 
and regard the (full) probe  $v= (v_{y} + iv_{x})/\sqrt{2} =[v(y)]^{MN}$
as a $\lq\lq$gauge field" with the transformation law, 
\begin{equation}
G(Z + iv)G^{\dag} = Z + i v^{G}.
\end{equation}
Remarkably, by a special choice of $G = e^{iS}$ with $S\sim O(v)$, 
the transformed field $v^{G} = v -[Z, S] +\cdots$, to $O(v)$, 
is expressed in multipoles of the magnetic field $b_{z}^{({\bf a)}} = \nabla  \times {\bf a}$ alone.
The field $v_{x}^{G} \in v^{G}$, being real, 
is naturally symmetric in orbitals $(M,N)$, $[v_{x}^{G}]^{MN} = [v_{x}^{G}]^{NM}$,
and is written to $O(v)$ as 
\begin{eqnarray}
[v_{x}^{G}]^{NN}\!\! &=& 0,
\nonumber\\
{[v_{x}^{G}]^{N, N+ 1}}\!\! &=&\! f^{N, N+1}_{p} u_{x} (y_{0})
= c^{1}_{N} L_{N}^{1}(\xi) du_{x} (y_{0}),
\nonumber\\
{[v_{x}^{G}]}^{N, N+ r}\!\! &\stackrel{r \ge 2}{=}&   2f^{N, N+r}_{p} u_{x} (y_{0})
\nonumber\\
&=& c_{N}^{r}\,  L_{N}^{r}\,(\xi)(d/\sqrt{2} )^{r-2} d^2 u_{x} (y_{0}),\ \ \ 
\label{vx_transG}
\end{eqnarray}
where $c^{1}_{N} = 1/\sqrt{2(N+1)}$ and $c_{N}^{r} \stackrel{r \ge 2}{=} \sqrt{N!/(N+r)!}$;
 $\xi \equiv {1\over{2}} p^2 = - {1\over{2}} d^2$  (with $d=\partial_{y_{0}}$) is a derivative acting
on $u_{x}(y_{0})$;
$du_{x}(y_{0}) \equiv - \gamma_{p}b_{z}(y_{0})$ is a magnetic probe 
with $b_{z}(y_{0}) \equiv -dv_{x}(y_{0}) =e\ell^2 b_{z}^{({\bf a})}(y_{0})$. 
See Appendix A, which summarizes the gauge transformation $G$ in a form suited for practical calculations.

At the same time $A_{0}(y)$ is transformed into 
$A_{0}^{G} \equiv GA_{0}(y)G^{\dag} = A_{0}(GyG^{\dag})$.
In components $[A_{0}^{G}]^{MN} = [A_{0}^{G}]^{NM}$, 
it is written to $O(d A_{0})$  as   
\begin{eqnarray}
[A_{0}^{G}]^{NN} \!\!
&=& A_{0}\big(y_{0} +  L_{N}(\xi) u_{x}(y_{0}) \big) ,
\nonumber\\
{[A_{0}^{G}]}^{N+1,N} \!\!\!
&=&  c^{1}_{N} \, dA_{0}(y_{0}) 
\big\{1 + {\textstyle {1\over{2}}}\, L_{N}^{1}(\xi) d u_{x}(y_{0}) \big\}. \ \ 
\end{eqnarray}
Here, for clarity, we retain only terms up to $O(d A_{0})$.  
See Appendix A for the full matrix form of $A_{0}^{G}$.

This transformation $G$ in the $\langle y|N \rangle$ basis is promoted~\cite{ks_edgeC_W,footnote_one} 
to the field $\psi^{m}$ and ${\cal H}^{mn}$ in the $(n,y_{0})$ space 
by setting
$\psi^{G}_{m} = {\cal G}^{mn} \psi_{n}$ and ${\cal H}^{G} ={\cal G}\, {\cal H}\, {\cal G}^{\dag}$ 
with 
\begin{equation}
{\cal G} = b' G b' + c G c + b G  b + c' G c',
\end{equation} 
where $(b' G b')^{mn} = b'_{m}G^{M-2,N-2}b'_{n}$, etc.;   ${\cal G} {\cal G}^{\dag} =1$. 
The transformed Hamiltonian then takes the form  
\begin{eqnarray}
H &=& \int dy_{0}\sum_{m, n}
(\psi^{G})^{\dag}_{m}(y_{0})\, ({\cal H}^{G})^{mn}\, \psi^{G}_{n}(y_{0}),
\label{H_GaugeT}\\
{\cal H}^{G}\!\!\! &=& {\cal H}_{0} + {\cal V}^{G}_{v} +  {\cal V}^{G}_{A},
\nonumber\\
{\cal V}^{G}_{v} 
&=&
\omega_{c} {1\over{\sqrt{2}}} \big( b\,  v_{x}^{G} c + c\, v_{x}^{G}b  + b' v_{x}^{G} c'  + c' v_{x}^{G} b' \big),
\nonumber\\ 
{\cal V}^{G}_{A} &=&- e \big( b A_{0}^{G} b + c A_{0}^{G}c + b' A_{0}^{G}b' + c' A_{0}^{G}c' \big),
\end{eqnarray}
with $({\cal H}_{0})^{mn} \equiv \epsilon_{n} \delta^{mn}$.
The diagonal terms $({\cal H}^{G})^{nn} \equiv {\cal E}_{n}(y_{0})$ represent the spectra of level $n$, 
diagonal to $O(a_{x})$, $O(A_{0})$ and $O(a_{x}\partial A_{0})$, 
\begin{eqnarray}
{\cal E}_{n}(y_{0}) &=& \epsilon_{n}[y_{0}]  + V_{n}[y_{0}],
\label{EnKone}
\\
\epsilon_{n}[y_{0}] &=&\epsilon_{n}-  e g^{nn}(\xi)\, \gamma_{p}A_{0}(y_{0})
\nonumber\\
&\approx& \epsilon_{n}   -e A_{0}(y_{0}) +  O(d^2A_{0}), 
\\
V_{n}[y_{0}] &=& \omega_{c} \kappa^{nn} (\xi) \,  du_{x}(y_{0}) 
\nonumber\\
&&
+  e \ell E_{y}(y_{0})\,  g^{nn}(\xi) u_{x}(y_{0}),
\label{HG_v_A}
\end{eqnarray}
with electric field $E_{y}(y_{0}) \equiv  -\partial_{y_{0}} A_{0}(y_{0})$.
Here  
\begin{eqnarray}
g^{nn}(\xi)
&=& b_{n}b_{n} L_{N-1}(\xi) + c_{n}c_{n} L_{N}(\xi) 
\nonumber\\ 
&& + b'_{n}b'_{n} L_{N-2}(\xi) + c'_{n}c'_{n} L_{N-1}(\xi),
 \label{gnn_xi}
\\
\kappa^{nn}(\xi) 
 &=& {b_{n}\, c_{n}\over{\sqrt{N}}}\, L^{1}_{N-1}(\xi) +  {b'_{n}\, c'_{n}\over{\sqrt{N-1}}}\,\, L^{1}_{N-2}(\xi) ,
\end{eqnarray}
act on $u_{x}(y_{0})$  with $\xi = -{1\over{2}}d^2$; 
$L_{N}(0) =1$ and $L_{N}^{1}(0) =N+1$; $g^{00}(\xi)=1$ and  $g^{11}(\xi)=1-(c_{1})^2 \xi$.

Off-diagonal portions of ${\cal V}^{G}_{v} + {\cal V}^{G}_{A}$, 
via further diagonalization,
yield corrections of the form  $(dA_{0})d^2 u_{x}$.
Such short-ranged terms lead to locally fluctuating current components, carrying no net amount, 
which are made practically invisible by other dominant components~\cite{ks_edgeC_W}. 
We therefore omit them and analyze ${\cal E}_{n}(y_{0})$ below.

These spectra ${\cal E}_{n}(y_{0})$ and their basic relation in Eq.~(\ref{ConjugatedSpec}) 
are essentially the same in form 
as those for monolayer graphene~\cite{ks_PNJ_G}, 
except that $g^{nn}_{p}$ and $\kappa^{nn}_{p}$ contain bilayer components $(b', c')$.
One can therefore immediately draw their consequences.

(i) From the $d u_{x}$ term in $V_{n}[y_{0}]$ one can read 
an orbital magnetic moment of a Dirac electron of level $n$, 
\begin{eqnarray}
 \hat{m}_{n} &=& e\ell^2  \omega_{c}\, \kappa^{nn}(0),
 \nonumber\\
\kappa^{nn}(0) &=&   b_{n}c_{n}\sqrt{N} + b'_{n}c'_{n} \sqrt{N-1}.
\label{MagM_n}
\end{eqnarray}
Actually
$\hat{m}_{n}$ takes a concise form 
if one calculates it from the $B$ dependence 
of the spectra $\epsilon_{n}$  in Eq.~(\ref{En_zeroth})  for zero bias $\mu \rightarrow0$
[by noting that $\delta g^2 = - (g^2/B) \delta B$ for $g \equiv \gamma_{1}/\omega_{c}$],
\begin{equation}
\hat{m}_{n}= - {\partial \epsilon_{n}\over{\partial B}} 
= - {1\over{2}} e\ell^2  \Big(1+{g^2 \over{\sqrt{D_{N}}}}\Big)\,  \epsilon_{n} \ \  {\rm for}\ N\ge 2;
\end{equation}  
Eq.~(\ref{MagM_n}) yields the  same result. 
In the $N \ge 2$ sectors, $\hat{m}_{n}$ change sign for $n=\pm N$ and 
grow like $\mp \sqrt{N}$ with $N$.
This sign change of  $\hat{m}_{\pm N}$ implies that,  as noted in Eq.~(\ref{eh-conj_jc}), 
positive- and negative-energy electrons undergo cyclotron motion  
in opposite directions to each other. 
The associated circulating currents $j_{n}^{\rm (c)}$ are diamagnetic $(\hat{m}_{N}^{K} < 0)$ 
for positive-energy electrons and are paramagnetic $(\hat{m}_{-N} > 0)$ 
for negative-energy ones.

In contrast, Eq.~(\ref{MagM_n})  reveals a weak and unusual magnetic response of 
PZM electrons $\ni (0_{-}^{K}, 1_{-}^{K}; 0_{+}^{K'}, 1_{+}^{K'})$, 
\begin{eqnarray}
\hat{m}_{n=0_{-}}^{K} &=&\hat{m}_{n=0_{+}}^{K'}  = 0,
\nonumber\\
\hat{m}_{n =1_{+}}^{K'} &=& -  \hat{m}_{n =1_{-}}^{K} 
=  e \ell^2 U_{z}\, (c_{1})^4/g^2 >0.
\label{OrbMag_PZM}
\end{eqnarray} 
In the $N=0$ sector Dirac electrons thus make no cyclotron motion.
In the $N=1$ sector, positive-energy electrons (of  $n=1_{+}^{K'}$) 
slowly orbit oppositely to other positive-energy ones, and 
the associated current $j^{\rm (c)}$ is paramagnetic with $\hat{m}_{1_{+}} >0$; see Eq.~(\ref{jc_one_Kp}).
Likewise the $n=1_{-}^{K}$ electrons orbit oppositely to other negative-energy ones, with a diamagnetic current. 
Note that Eq.~(\ref{OrbMag_PZM}) also follows from the $B$ dependence of the spectra 
$\epsilon_{0_{+}}^{K'} = {1\over{2}} U_{z}$, 
$\epsilon_{1_{+}}^{K'} \approx {1\over{2}} U_{z} (g^2-1)/(g^2+1)$, etc.
It is now clear that the paramagnetic nature of the $n=1_{+}^{K'}$ electrons, e.g., 
derives from their pseudo-zero spectrum $\epsilon_{1_{+}}^{K'}$ 
which gets lower as $B$ is increased.
Actually, the paramagnetism intrinsic to the PZM sector was noticed earlier 
in thermodynamic calculations~\cite{KA}.

To derive a real-space response from $V_{n}[y_{0}]$
one has to relate $u_{x}(y_{0})$ to the real-space field $v_{x}(y)$, using, 
e.g., 
$[v_{x}(y)]^{NN} = L_{N}(\xi)u_{x}(y_{0}) = \int\! dy\, v_{x} (y)  |\phi_{N}(y-y_{0})|^2$
and the formula 
$L_{N-1}^{1}(x) = \sum_{n=0}^{N-1} L_{n}(x)$.
The $\kappa^{nn}\, du_{x}$ term then leads to 
the magnetization density of level $n$,
\begin{eqnarray} 
M^{\rm (c)}_{z;n }(y) &=& {e\ell^{2} \omega_{c}\over{L_{x}}} \int dy_{0}\, 
{\cal K}_{n}(y-y_{0})\, \hat{\rho}_{n}(y_{0}),
\label{Mzn_y}
\end{eqnarray}
where $\hat{\rho}_{n}(y_{0}) = \hat{\psi}_{n}^{\dag}(y_{0}) \hat{\psi}_{n}(y_{0})$ 
is the electron density in the $(n, y_{0})$ space and 
\begin{eqnarray}
{\cal K}_{n}(y) 
 &=&  {b_{n}c_{n}\over{\sqrt{N}}}\, {\cal D}_{N}(y)
+ {b'_{n}c'_{n}\over{\sqrt{N-1}}}{\cal D}_{N-1}(y), 
\nonumber\\
 {\cal D}_{N}(y) 
  &=& |\phi_{0}(y)|^2 + |\phi_{1}(y)|^2 + \cdots+ |\phi_{N-1}(y)|^2. \
\end{eqnarray}
In the PZM sector,  ${\cal K}_{0_{\pm}}(y) =0$ and
\begin{equation}
{\cal K}^{K'}_{1_{+}}(y) = - {\cal K}^{K}_{1_{-}}(y) =  2 \mu \{(c_{1})^4/g^2\} |\phi_{0}(y)|^2  > 0
\end{equation}
while ${\cal K}_{n}(y) < 0$ and ${\cal K}_{- n}(y) > 0$ for $n \ge 2$.

Let us now suppose that the static potential $A_{0}(y)$ is applied 
so that some lower levels are confined in a certain domain of a sample. 
Each filled level $n$ is then characterized by a filled domain 
$\{y_{0};  y_{0;n}^{-} \le y_{0} \le y_{0;n}^{+}\}$ in the center space $\{y_{0}\}$
and the endpoints $y_{0;n}^{\pm}$ are fixed from the spectrum 
$\epsilon_{n}[y_{0;n}^{\pm}]  \approx \epsilon_{n}  - e A_{0}(y_{0;n}^{\pm}) = \epsilon_{\rm F}$
for a given value of the Fermi energy $\epsilon_{\rm F}$.
Each filled domain has a constant density 
$\langle \hat{\rho}_{n}(y_{0}) \rangle/L_{x} =1/(2\pi \ell^2) \equiv \bar{\rho}$. 
 Taking an expectation value of $M^{\rm (c)}_{z;n }(y)$ for a filled level, 
\begin{equation}
\langle M^{\rm (c)}_{z;n }(y)\rangle = {e\,  \omega_{c}\over{2\pi}} \int_{y_{0;n}^{-}}^{y_{0;n}^{+}} dy_{0}\, 
{\cal K}_{n}(y-y_{0}),
\end{equation}
yields a real-space magnetization density. 
The associated {\it circulating} current 
$j_{n}^{\rm (c)} = (\nabla \times {\bf M}^{\rm (c)})_{x} = \partial_{y} M^{\rm (c)}_{z;n}$ 
then reveals a distribution  
\begin{equation}
\langle j^{\rm (c)}_{n}(y)\rangle
= - {e \omega_{c}\over{2\pi}}
\Big\{  {\cal K}_{n}(y-y_{0;n}^{+}) -  {\cal K}_{n}(y-y_{0;n}^{-})\Big\}.\ \ \ 
\label{jc_density}
\end{equation}
Naturally  $\langle j^{\rm (c)}_{n}(y)\rangle$ cancels out in a densely populated domain 
and survives only along the periphery $y \sim y_{0;n}^{\pm}$. 
Near the upper edge, in particular, 
the PZM electron levels $(0_{+}^{K'},1_{+}^{K'})$ show the current profile
\begin{eqnarray}
\langle j^{\rm (c)}_{0_{+}}(y) \rangle &=& 0,  
\nonumber\\
\langle j^{\rm (c)}_{1_{+}}(y)\rangle
&=& - {e\over{2\pi}} {(c_{1})^4\over{g^2}}\, U_{z}  |\phi_{0}(y- y^{+}_{0;1})|^2 <0
\label{jc_one_Kp}
\end{eqnarray}
while $\langle j^{\rm (c)}_{n}(y \sim y_{0;n}^{+})\rangle >0$ in  higher  levels $n\ge 2$.
The amount of current per (upper) edge naturally reflects the magnetization density in the filled domain,
\begin{equation} 
J_{n}^{{\rm (c)};+} =   \int dy\, \langle j_{n}^{\rm (c)}(y) \rangle 
= - {e\omega_{c}\over{2 \pi}} \ \kappa^{nn} (0)= -\bar{\rho}\,\hat{m}_{n}.
\label{Jn_c}
\end{equation}

(ii)~On the other hand, the $O(v_{x} A_{0})$ response in $V_{n}[y_{0}]$ 
detects the drift (or Hall) current $j^{\rm (d)}$, driven by a local field  $E_{y}(y_{0})$, 
with a real-space distribution
\begin{equation}
\langle j^{\rm (d)}_{n}(y) \rangle = -{e^2\over{2\pi}} 
 \int_{y_{0;n}^{-}}^{y_{0;n}^{+}} dy_{0}\,  {\cal R}_{n} (y-y_{0})\, E_{y}(y_{0}),
\end{equation}
where $ {\cal R}_{n}(y)$ is given by 
$g^{nn}(\xi)$ in Eq.~(\ref{gnn_xi}) with replacement $L_{N-1}(\xi) \rightarrow  |\phi_{N-1}(y)|^2$, etc.; 
${\cal R}_{0}(y) = |\phi_{0}(y)|^2$ and ${\cal R}_{n}(y) =\{ 1+ O(\partial_{y}^2)\}{\cal R}_{0}(y)$.
Away from the edges $(y_{0;n}^{-}\ll y \ll y_{0;n}^{+})$,
\begin{equation}
\langle j_{n}^{\rm (d)}(y)  \rangle 
= -(e^2/2 \pi) E_{y}(y) + O(\partial E_{y}) 
\label{jby_Wlocal}
\end{equation}
locally flows with a unit conductance $\sigma_{xy} = -e^2/(2\pi \hbar)$ per level.
As noted in  Eq.~(\ref{jhall_KKp}), 
$j_{n}^{\rm (d)}(y)$ flows in the same direction  for all $\{n\}$ in both valleys, 
and is  paramagnetic, 
with $\langle j_{n}^{\rm (d)}(y)  \rangle  <0$ at $y\sim y_{0;n}^{+}$ (where $E_{y}>0$).

A practical platform to observe these currents   $(j^{\rm (c)}_{n}, j^{\rm (d)}_{n})$ 
is a $p$-$n$ junction set up in a gated sample.
In graphene they are observed~\cite{UKBL} as two species of equilibrium current, 
that alternate direction along the junction and spatially shift 
as $\epsilon_{\rm F}$ is varied. 
When a $p$-$n$ junction is formed in bilayer graphene, 
the PZM levels get orbitally mixed in each valley across the junction, in which $E_{y}(y)\not=0$, 
and their splitting $\Delta e_{10} = |e_{1}-e_{0}| \approx 2\mu (c_{1}/g)^2$ is enhanced,
 \begin{equation}
 \Delta e_{10}  \rightarrow  \sqrt{ (\Delta e_{10})^2 + 2\{c_{1})^2 \{eE_{y}(y_{0})/\omega_{c}\}^2 }.
\end{equation}
Still the mixing angle $\theta (y_{0})$, unlike the one in Eq.~(\ref{MixingAngle}), 
stays $|\theta|< \pi/2$ and the associated Berry phase
becomes trivial $\Delta \theta =0$, as readily verified.
Apart from this character of the PZM sector, 
current distributions around a $p$-$n$ junction will share essentially the same features 
as those in monolayer graphene, examined earlier~\cite{ks_PNJ_G}.
Accordingly, we skip further discussion in this direction 
and turn to the valley-contrasting junction.

\section{Current associated with the  kink states }

In this section we turn off  $A_{0} \rightarrow 0$ 
and explore observable features of the PZM states formed
along the kink of interlayer bias $U_{z}(y) =2\omega_{c} \mu(y)$ in equilibrium.
Let us now couple  $(a_{x}, A_{0})$ to the PZM Hamiltonian ${\cal H}_{\rm PZ}$ in Eq.~(\ref{H_PZM}),
and gauge transform it,  as in Eq.~(\ref{H_GaugeT}), to have the Hamiltonian
$H = \int dy_{0}
(\psi^{G})^{\dag}_{m}(y_{0})\, ({\cal H}^{G})^{mn}\, \psi^{G}_{n}(y_{0})$
with 
${\cal H}^{G} = {\cal H}_{0} + {\cal V}^{G}_{v} +  {\cal V}^{G}_{\mu}$,
where 
\begin{equation}
{\cal V}^{G}_{\mu} = -\omega_{c} 
\big\{ b\, \mu^{G} b + c\,  \mu^{G} c - b'\,  \mu^{G} b' - c'\, \mu^{G} c' \big\}
\label{V_mu_G}
\end{equation}
denotes the transformed bias with
$\mu^{G}(y) \equiv G \mu(y) G^{\dag}$.  
We expand $[\mu^{G}(y)]^{MN}$ in multipoles of $[\mu(y)]^{00} = m_{0}(y_{0})$ up to $O(d^2 m_{0})$,
diagonalize, as before, the spectra 
$(\hat{\cal H}^{G})^{mn} = (U{\cal H}^{G}U^{\dag})^{mn}$ with $\hat{\psi} =  U\,  \psi^{G}$ 
and extract the $O(v_{x})$ response from them. 
See Appendix B,  where this step of calculation is elaborated.

The resulting PZM spectra 
$(\hat{\cal H}^{G})^{nn} \equiv {\cal E}_{n}^{K}(y_{0}) = \omega_{c}\, \hat{e}^{G}_{n}[y_{0}]$ in valley $K$,
diagonal to $O(v_{x})$, 
are written as 
\begin{equation}
\hat{e}_{n}^{G}[y_{0}] = \hat{e}_{n}^{\rm (c)}[y_{0}] +  \hat{e}_{n}^{\rm (d)}[y_{0}] \ \ \ \ (n=0,1).
\label{hat_eG_n}
\end{equation}
Here $\hat{e}_{n}^{\rm (c)}[y_{0}]$ refer to  a magnetic response, 
\begin{eqnarray}
\hat{e}_{n}^{\rm (c)}{[y_{0}]} &=& \kappa_{n} (y_{0}, \xi) du_{x}(y_{0}), 
\\
\kappa_{n}(y_{0}, \xi)  
&\stackrel{n=1/0}{=}& (c_{1}/g)^2 \{ {\cal K}(y_{0}, \xi) \pm {\cal K}_{D}  (y_{0}, \xi) \}, 
\label{hat_eG_circling}
\end{eqnarray}
with  $d u_{x}(y_{0}) = - \gamma_{p}b_{z}(y_{0})$ and 
\begin{eqnarray}
 {\cal K} (y_{0}, \xi) 
 &=&   -(c_{1})^2 Q +  {c_{1}\over{4\sqrt{2}}} dR\, \{ 5 +  L_{1}(\xi) \},
\nonumber\\
 {\cal K}_{D}(y_{0}, \xi) &=& c(\theta) \Big\{  {\cal K} (y_{0}, \xi) - {dR\over{2\sqrt{2}\, c_{1}}} \Big\} 
+  {1\over{2}} s(\theta) R, \ \ \ \ 
 \end{eqnarray}
where $Q = Q(y_{0})$, $R = R(y_{0})$, 
$c(\theta) \equiv \cos \theta(y_{0}) = Q/\sqrt{D}$ and $s(\theta) \equiv \sin \theta (y_{0})= R/\sqrt{D}$.

 For a given $\epsilon_{\rm F} >0$ there are two active channels 
$\{y_{0;n}^{-} \le y_{0} \le y_{0;n}^{+}\}|_{n=0,1}$ per valley and spin,
as seen from Fig.~2(b).
The PZM levels $(\hat{0}^{K}, \hat{1}^{K})$ over each filled domain 
acquire an orbital magnetization density  and a current density of the  form  
\begin{eqnarray}
\langle M^{\rm (c)}_{z;n}(y) \rangle &=&  {e \omega_{c}\over{2\pi}} \int_{y_{0;n}^{-}}^{{y_{0;n}^{+}}} dy_{0}\, 
I^{\rm (c)}_{n}(y,y_{0}), 
\\
I^{\rm (c)}_{n}(y,y_{0})\!\! &=& \!\!
T_{0}(y-y_{0})\,  \kappa_{n}(y_{0}, 0)  + T_{1,0}(y-y_{0})  \kappa^{(\xi)}_{n}(y_{0}),
\nonumber\\
\langle j^{\rm (c)}_{n}(y) \rangle &=&  
\partial_{y} \langle M^{\rm (c)}_{z;n}(y) \rangle,
\end{eqnarray}
where $\kappa^{(\xi)}_{n =1/0}(y_{0}) \equiv  (c_{1}/g)^2 (c_{1}/4\sqrt{2})\{1\pm c(\theta)\}dR$
refers to a factor multiplying $L_{1}(\xi)$ in $ \kappa_{n}(y_{0}, \xi)$
and 
\begin{equation}
T_{0} (y) =  |\phi_{0}(y)|^2,\ \ 
T_{1,0} (y) = |\phi_{1}(y)|^2 - |\phi_{0}(y)|^2.
\end{equation}
As inferred from Fig.~2 and Eq.~(\ref{jc_density}), 
the circulating current  $j^{\rm (c)}(y) = j^{\rm (c)}_{0}(y) + j^{\rm (c)}_{1}(y)$ 
displays a characteristic peak profile 
near the upper edge $y\sim y_{0;0}^{+}$ of level $\hat{0}^{K}$,  
\begin{equation}
\langle j^{\rm (c)}(y) \rangle \approx - {e \omega_{c}\over{2\pi}}\, T_{0}(y-y_{0;0}^{+})\,  \kappa_{0} (y_{0;0}^{+}, 0)
+ O(d \kappa_{0}),  
\label{jc_at_edge}
\end{equation}
which approaches $\langle j^{\rm (c)}_{1_{+}}(y)\rangle$ in Eq.~(\ref{jc_one_Kp}) 
as level $\hat{0}^{K} \rightarrow 1_{+}^{K}$.

On the other hand, $\hat{e}_{n}^{\rm (d)}[y_{0}]$  involves an electric response 
probing the drift current  $j^{\rm (d)}_{n}$, 
\begin{eqnarray}
\hat{e}_{1}^{\rm (d)}[y_{0}] &=& \hat{e}_{1}(y_{0} + u_{x}) + \Lambda^{+} (y_{0}) \{L_{1}(\xi) -1\}\,u_{x},
\nonumber\\
\hat{e}_{0}^{\rm (d)}[y_{0}] &=& \hat{e}_{0}(y_{0} + u_{x}) + \Lambda^{-} (y_{0})  \{L_{1}(\xi) -1\}\,u_{x},
\nonumber\\
\Lambda^{\pm}(y_{0})\!
&=&\! {(c_{1})^3\over{\sqrt{2}\, g^2}} R\Big \{1 \pm  {1\over{\sqrt{D}}} (Q+ {1\over{\sqrt{2}c_{1}}}dR) \Big\},\ \ \ \ 
\label{jd_paramag}
\end{eqnarray}
where 
$\{L_{1}(\xi)-1\}u_{x} = {1\over{2}}d^2 u_{x}$. 
Note that the lowest multipole $u_{x} = u_{x}(y_{0})$ is accommodated 
into the argument of the spectra $\hat{e}_{n}(y_{0})$ in Eq.~(\ref{Kink_Spectra}); 
this feature derives from electromagnetic gauge invariance.
This structure generally tells us that electrons driven by the gradient of the spectrum 
(or by the group velocity) carry a paramagnetic current. 
This is clear from the sign of current $\langle j_{n} \rangle \propto - d{\cal E}_{n}(y_{0})  <0$ 
at the upper edge where $d{\cal E}_{n}(y_{0})>0$.

From $\hat{e}_{n}^{\rm (d)}[y_{0}]$ follows the real-space distribution of the drift current 
over the same filled domain,
\begin{eqnarray}
\langle j^{\rm (d)}_{n}(y) \rangle 
&=& - {e\, \omega_{c}\over{2\pi}} \int_{y_{0;n}^{-}}^{y_{0;n}^{+}} \!\! dy_{0}\, I^{\rm (d)}_{n}(y,y_{0}),
\\
I^{\rm (d)}_{1} (y, y_{0})\!\!
&=&\! T_{0}(y\! -y_{0})  d\hat{e}_{1}(y_{0}) + T_{1,0}(y\! -y_{0}) \Lambda^{+}(y_{0}), 
\nonumber\\
I^{\rm (d)}_{0} (y, y_{0}) \!\!
&=&\! T_{0}(y\! -y_{0})   d\hat{e}_{0}(y_{0}) + T_{1,0}(y\! -y_{0}) \Lambda^{-}(y_{0}).\ \ \ \ \  
\label{jd_Kink}
\end{eqnarray}
In valley $K$ alone the net amount of this current
\begin{eqnarray}
&&\int_{-\infty}^{\infty} dy \langle  j^{(d)}_{n}(y) \rangle 
= - {e\, \omega_{c}\over{2\pi}} \int_{y_{0;n}^{-}}^{{y_{0;n}^{+}}} dy_{0}\,  d\hat{e}_{n}(y_{0}) 
\nonumber\\
&&= - {e\, \omega_{c}\over{2\pi}} \{ \hat{e}_{n}(y_{0;n}^{+}) -   \hat{e}_{n}(y_{0;n}^{-}) \}
= - {e\over{2\pi}}\, \epsilon_{\rm F}\ 
\label{NetAmount_jd}
\end{eqnarray}
is simply fixed by the energy difference between the two ends $y_{0} =y_{0;n}^{\pm}$, 
or by the Fermi energy $\epsilon_{\rm F} \lesssim U_{z}/2$. 
In another valley $K'$ with spectra 
$\hat{e}_{n}^{{\rm (d)};K'}[y_{0}] = -\hat{e}_{n}^{{\rm (d)};K}[y_{0}]$, 
as seen from Fig.~2(b),
$j_{n}^{\rm (d)}$ flows in the opposite direction with a net amount $(e/2\pi)\epsilon_{\rm F}$ per level.
This confirms that the drift current $j^{(d)}_{n}(y) \propto O(\mu)$ is  
a valley Hall current driven by the bias field 
$e E_{y}^{\rm eff} = \omega_{c}d\hat{e}_{n}(y_{0})$,
with a unit conductance $\sigma_{xy} = \mp e^2/(2\pi\hbar)$ per level in valley $K/K'$.

In equilibrium the total current 
$j_{n}^{{\rm (d)};K + K'}(y) = j_{n}^{{\rm (d)};K}(y) + j_{n}^{{\rm (d)};K'}(y)$ [with $n=(0,1)$] 
thus simply circulates along the junction as a persistent current.
Figure~2(b) shows four channels of active edge modes (per spin) 
that propagate around two focuses $(y_{0;1}^{-},  y_{0;0}^{-})$. 
[To imagine how they circulate, let us suppose that kink states terminate at free ends for $x \rightarrow \pm \infty$. 
 Valley current channels will then merge together there and change their valleys and directions simultaneously.]
It is clear now that the total drift current 
$j^{{\rm (d)};K+K'}_{n}(y)$ 
is also written as a circulating current associated 
with a $\lq\lq$drift" orbital magnetization, denoted as $M_{z;n}^{\rm (d)}(y)$,
\begin{equation}
\langle j^{{\rm (d)};K+K'}_{n}(y) \rangle 
= \partial_{y}  \langle M_{z;n}^{\rm (d)}(y) \rangle.
\end{equation}  
The operator 
$M_{z;n}^{\rm (d)}(y) =\int_{-\infty}^{y} dy\, j^{{\rm (d)}; K+K'}_{n}(y)$ 
is constructed from $j^{\rm (d)}_{n}(y)$ in Eq.~(\ref{jd_Kink}) 
by replacing the kernels $T_{0}$ and $T_{1,0}$ with their integrals $\int_{-\infty}^{y} dy\,  T_{0} (y-y_{0})$, etc.

Finally we refer to a charge probe. 
When $A_{0}(y)$ is retained to detect charge distributions, 
the spectra ${\cal E}_{n} =\omega_{c}\, \hat{e}^{G}_{n}$ acquire the following response 
\begin{equation}
\delta_{A}{\cal E}_{1} =-  A^{00} 
-  {c_{1}\over{\sqrt{2}}}    {R\over{\sqrt{D}}}  dA^{00}\,  
- {1\over{4}} (c_{1})^2 (1+ {Q\over{\sqrt{D}}} ) d^2A^{00}, 
\end{equation}
with $A^{00} \equiv e\gamma_{p}A_{0}(y_{0})$;  
for $\delta_{A}{\cal E}_{0}$ set $(Q, R) \rightarrow (-Q,-R)$. 
They are used to calculate the charge density $q(y) = q_{0}(y) + q_{1}(y)$ 
shown in Fig.~3.

%%%%%%%%%%%%%%% Figure3 %%%%%%%%%
\begin{figure*}[tbp]
\begin{center}
\includegraphics[scale=1.2]{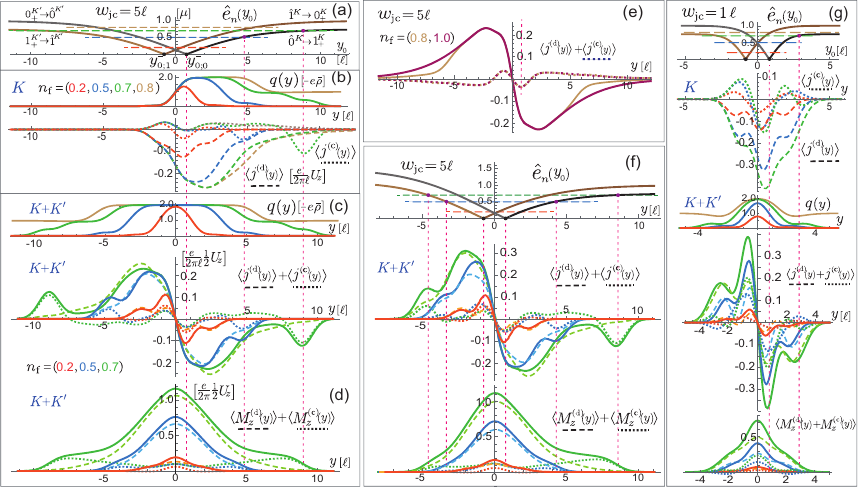}
\end{center}
\caption{
Electron kink states formed along a valley-contrasting junction. 
(a)~PZM spectra above the charge-neutral ground state under a bias $\mu(y)$ with $w_{\rm jc} =5 \ell$. 
Charge density $q(y)$ and current density  
$j_{x}(y) = j^{\rm (c)}(y) +  j^{\rm (d)}(y)$ 
that develop (b) in valley $K$ alone and (c) in full valleys $K+K'$ 
at filling $n_{\rm f}=(0.2,0.5, 0.7, 0.8)$. 
The current $j_{x}$ flows in one way on each side of the junction, 
showing two peak profiles associated with $j^{\rm (d)}$ (dashed curves) 
and $j^{\rm (c)}$ (dotted curves).
(d)~Orbital magnetization density  
$M_{z}(y)  =  M_{z}^{\rm (d)}(y) + M_{z}^{\rm (c)}(y)$
 in full valleys $K+K'$. 
The drift component $\langle M_{z}^{\rm (d)}(y) \rangle$ grows steadily with the kink state.  
In contrast, $\langle M_{z}^{\rm (c)}(y)  \rangle$ slowly grows toward the edges $y \sim \pm y_{0;0}^{+}$
and, upon vanishing, generates peak profiles of $j^{\rm (c)}$ 
that move away with increasing $n_{\rm f}$, 
as indicated by some vertical dotted lines. 
(e)~The accumulated current survives even when the PZM levels are filled up.
(f)~Kink states formed under a left-right asymmetric bias $\mu(y)$.
(g)~Under a sharp kink bias $\mu(y)$ with $w_{\rm jc}= 1 \ell$, the emerging electron state and 
associated currents are tightly localized at the junction. 
}
\end{figure*}
%%%%%%%%%%%%%%%%%%%%%%%%%%%%%%%%%%%%%%

For numerical simulations we first adopt an interlayer bias 
$\mu (y) = - \mu\, {\rm tanh} (y / w_{\rm jc})$ 
with $w_{\rm jc}= 5 \ell$ and examine the electron states formed along the junction $y \sim 0$.
We set $\epsilon_{\rm F} = \omega_{c} \mu\,   n_{\rm f}\ = {1\over{2}}\,  U_{z}  n_{\rm f} >0$ and 
use $n_{\rm f}$ to specify their filling.

Figure 3(a) shows the electron portion of the PZM spectra 
$\{ \hat{e}^{K'}_{n}(y_{0}), \hat{e}^{K}_{n}(y_{0}) \}|_{n= 0,1}$.
In valley $K$ alone, the electrons fill the spectra $\hat{e}^{K}_{n}(y_{0})$ with increasing $n_{\rm f}$, 
starting from the $\epsilon=0$ positions $(y_{0;0}^{-}, y_{0;1}^{-})$ in the $y_{0}$ space.
These positions are about 2$\ell$ apart, as noted in Eq.~(\ref{Focuses}), 
and the associated wave functions 
$\phi_{0}(y-y_{0;n})$ [of spread $\sim O(4\ell)$] spatially overlap each other.
Accordingly, in real $y$ space, the $n=(0,1)$ channels of kink states emerge as a single state.
This is clearly seen from the charge and current densities
$q(y)$ and $j_{x}(y) = j^{\rm (d)}(y) +  j^{\rm (c)}(y)$, 
shown in Fig.~3(b), that develop in valley $K$
with increasing filling $n_{\rm f} = (0.2, 0.5, 0.7, 0.8)$.
[Each density here refers to a sum of $n= (0,1)$ components (per spin).] 
Clearly visible in the current distributions are two peak profiles of the same sign,
showing that $j^{\rm (d)}(y)$ (dashed curves) and $j^{\rm (c)}(y)$ (dotted curves) flow in the same direction.

In full valleys $K+K'$, as shown in Fig.~3(c), a single real-space electron state develops monotonically 
around $y = 0$ and the charge density $q(y)$ extends a flat central portion 
with increasing filling $n_{\rm f}$. 
In this kink state the $K$- and $K'$-valley contents of the associated charge and current
 spatially overlap in the central region $|y| \lesssim 3 \ell$ while they stay distinct in wide outer regions.
 The valley contrast across a junction thus becomes more pronounced 
 with increasing magnetic field $B$ and filling $n_{\rm f}$, 
 as observed~\cite{LWMZ,LZYZ} in experiment.

It is enlightening to examine the current $j_{x}(y)$ in full valleys $K+K'$
through the associated orbital magnetization  
$M_{z}(y) = M_{z}^{\rm (d)}(y) + M_{z}^{\rm (c)}(y)$, 
depicted in Fig.~3(d).
From $\langle M_{z}(y) \rangle$ one can read  
the amount of current $\langle j_{x}(y) \rangle =\partial_{y} \langle M_{z}(y) \rangle$ 
in a given interval in real $y$ space;
in particular, $\langle M_{z}(0) \rangle$ tells us the amount of current flowing on each side $y<0$ or  $y>0$.

In both $\langle M_{z}(y)  \rangle$ and $\langle j_{x}(y) \rangle$, 
the drift components $\propto O(U_{z} n_{\rm f})$ dominate over 
the circulating ones $\propto O(U_{z} n_{\rm f} /g^2)$.
The drift magnetization $\langle M_{z}^{\rm (d)}(y)  \rangle$
steadily grows in height and width with filling $n_{\rm f}$, 
and leads to a broad and solid drift current $\langle j^{\rm (d)}(y)\rangle$ circulating along the junction.

Note that $\langle M_{z}^{\rm (d)}(0) \rangle \gg \langle M_{z}^{\rm (c)}(0) \rangle >0$.
The $(+)$ signs imply that  they both are paramagnetic, so are the associated currents.
The circulating current $j^{\rm (c)}(y)$ therefore flows, in net amount, 
in the same direction as $j^{\rm (d)}(y)$ on each side $y<0$ or  $y>0$. 
The magnetization $\langle M_{z}^{\rm (c)}(y) \rangle$ locally exhibits a small rise at the center $y = 0$ 
and slowly grows toward the edges $y\sim \pm y_{0;0}^{+}$, and then vanishes rapidly. 
The current $\langle j^{\rm (c)}(y)\rangle$ thereby acquires 
a small peak profile at  $y \sim \pm y_{0;0}^{-}$ and 
a prominent one, noted in Eq.~(\ref{jc_at_edge}), at  $y\sim \pm y_{0;0}^{+}$
while it reverses polarity and softens between them. 
The peaks at $y\sim \pm y_{0;0}^{+}$  move outward with increasing filling 
and eventually disappear when the $(1_{+}^{K}, 1_{+}^{K'})$ levels are filled (for $n_{\rm f} \gtrsim 0.75$).
In the figure dotted vertical lines refer to some such moving endpoints $y_{0;0}^{+}$.

When  the PZM levels are filled up (with $n_{\rm f} \ge 1$), 
the charge density attains a uniform distribution 
and the isolated kink state disappears.
Still the accumulated current $j^{\rm (d)}(y) +  j^{\rm (c)}(y)$ and orbital magnetization $M_{z}(y)$
survive over the junction, as depicted in Fig.~3(e).

Figure~3(f) shows the case of a left-right  asymmetric bias constructed from $\mu(y)$ 
by shifting and scaling it up slightly 
so that $\mu(y \ll 0) =  1.5 \mu$ and $\mu(y \gg 0)= - \mu$.
Clearly the orbital magnetization and associated current within the kink state exhibit such features 
as readily expected from the bias profile $\mu(y)$ 
and the filling $n_{\rm f}$ on each side of the junction.

On the other hand, under a very sharp kink bias with $w_{\rm jc} =1\ell$, 
the electron state is tightly localized at the junction, as shown in Fig.~3(g). 
The two species of current
$\langle j^{\rm (d)}(y)\rangle$ and $\langle j^{\rm (c)}(y)\rangle$ 
almost overlap spatially  and circulate in an accelerated manner along the junction.

So far we have examined the kink state formed in the electron regime  with  $\epsilon_{\rm F} >0$.
When $\epsilon_{\rm F} <0$, the kink state is realized as a hole state.
The absence of a negative-energy state $(-n, y_{0})$ is  
identified as a hole state with energy
${\cal E}_{n}^{\rm hole}\stackrel{n>0}{=} -{\cal E}_{-n} >0$ (per valley and with spin reversed), 
which, in view of Eq.~(\ref{ConjugatedSpec}), 
equals the energy of the $(n,y_{0})$ electron state in another valley, 
e.g., 
${\cal E}_{n}^{{\rm hole};K} \stackrel{n>0}{=}  {\cal E}^{K'}_{n}|_{-A_{0}}$; 
clearly, holes have positive charge $e>0$ and drift or circulate oppositely to electrons.
Thus ${\cal E}_{n}^{{\rm hole};K} + {\cal E}_{n}^{{\rm hole};K'} 
= {\cal E}_{n}^{K} + {\cal E}_{n}^{K'}$ for $A_{0}=0$.
This shows that the hole kink state realized for $\epsilon_{\rm F} <0$ in equilibrium
 shares  essentially the same current response 
as the electron kink state at $\epsilon_{\rm F} =|\epsilon_{\rm F}| >0$.

Finally it will be worth mentioning what happens when the polarity of gating 
is reversed, $\mu (y) \rightarrow -\mu (y)$. 
The answer comes from unitary equivalence in  Eq.~(\ref{unitary_equiv}). 
One simply has to interchange valley labels $K \leftrightarrow K'$ in all the figures. 
The valley currents and orbital magnetizations associated with the kink state thus remain paramagnetic, 
with only valley labels interchanged.

\section{Summary and discussion}

Characteristic of bilayer graphene in a magnetic field is an octet of PZM levels, forming the lowest Landau level.
Under an interlayer bias they split in valleys $(K,K')$ with a tunable valley band gap. 
One can form, by reversing the sign of the bias locally by gating, 
a line junction that contrasts the valley-polarized PZM electrons in adjacent domains on a sample.
A small portion of PZM levels lying inside the bulk band gap then supports low-energy electron (or hole) states 
localized along the junction.
These channels of kink states spatially appear as a single quasi-1D state 
since they all start to develop around the junction center,
with their $K$- and $K'$-valley contents spatially overlapping with spread of $O(\ell)$, 
as is clear from Fig.~3.  
Beyond this central portion the kink state contrasts distinct valleys on both sides;
the valley contrast becomes more pronounced with increasing electron filling $n_{\rm f}$ and magnetic field $B$.

The (spatially-single) kink state accommodates two species of current,  $j^{\rm (d)}$ and $j^{\rm (c)}$,
that flow in one-way on each side  or circulate along the junction in equilibrium. 
The dominant one, the drift current $j^{\rm (d)}$,  is essentially a Hall current driven by the bias gradient 
and is intrinsically paramagnetic, as remarked in relation to Eq.~(\ref{jd_paramag}).
The kink state thus provides a platform for the valley Hall effect~\cite{XYN} and 
the full-valley current $j^{{\rm (d)};K+K'}$ serves as a circulating valley Hall current.
The circulating current $j^{\rm (c)}$, associated with cyclotron motion, 
flows in essentially the same direction as $j^{\rm (d)}$,
as a result of an unusual magnetic response of the PZM electrons, 
as noted in Eqs.~(\ref{OrbMag_PZM}) and (\ref{jc_one_Kp}).

 %%%%%%%%%%%%%%% Figure4 %%%%%%%%%
\begin{figure}[tbp]
\begin{center}
\includegraphics[scale=0.425]{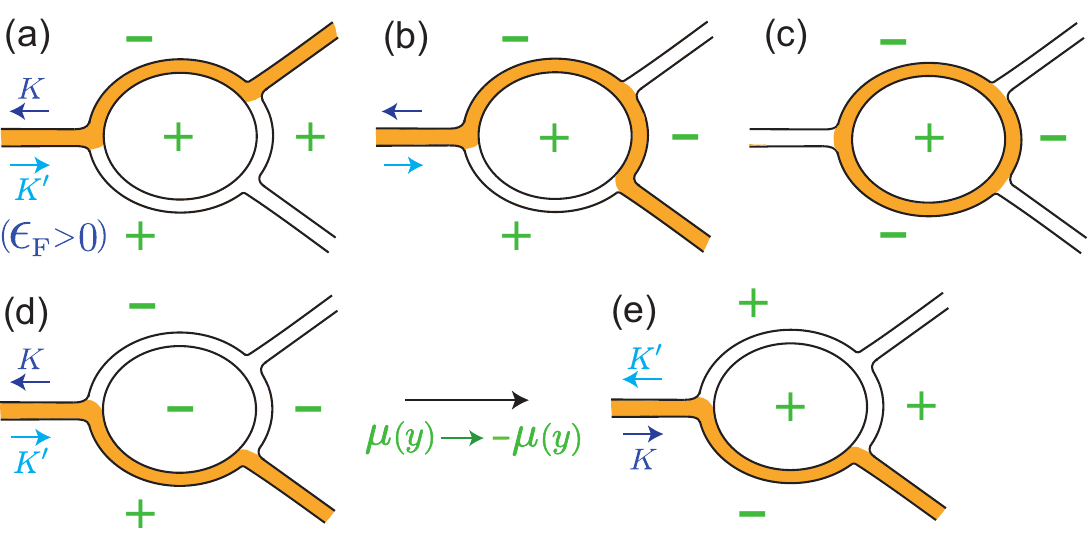}
\end{center}
\caption{
Orbital magnetization channels along a network of tunable valley-contrasting junctions in bilayer graphene in equilibrium.
It is possible, by the choice of bias patterns, to configure a variety of magnetization channels and alter them; 
even a closed channel can be set up. 
The quasi-1D channel of magnetization is always paramagnetic and, 
when the polarity of gating (indicated by the sign $\pm$) is reversed, 
only the valley labels of the associated currents are interchanged, $K \leftrightarrow K'$. 
}
\end{figure}
%%%%%%%%%%%%%%%%%%%%%%%%%%%%%%%%%%%%%%

 The current $j^{\rm (d)} + j^{\rm (c)}$ is a persistent current and 
 spatially circulates within the kink state in equilibrium, 
 creating a channel of orbital magnetization localized along the junction.  
 This magnetization channel will persist as long as the driving force of the kink bias 
 remains effective against disorder.
Figure 4 depicts schematically such a network of narrow orbital-magnetization channels
set up on a bilayer sample divided into four gated regions, 
in each of which the polarity of interlayer bias 
$\pm U_{z} =\pm 2\omega_{c} \mu$ is tunable independently.  
One can configure, by adjusting bias patterns, a variety of orbital magnetization channels and control them.
Even a closed channel can be induced, as in Fig.~4(c).
Current transport through a network of kink states was actively studied~\cite{QJNM,LWMZ,LZYZ,CZLQ}. 
In the present network the carrier of valley signals is a narrow channel of orbital magnetization 
present in equilibrium, with no injected current.  
Signals are detected by measuring magnetization locally at possible ends of the network.
Such electric control of orbital magnetization channels and associated valley currents 
will deserve serous attention in developing valleytronics applications.  

In the absence of a magnetic field, $B = 0$, chiral edge modes 
driven by the kink interlayer bias~\cite{MBM,QJNM,ZPFP}
carry the drift current $j^{\rm (d)}$. 
It is shown quite generally for $B=0$ in the present bilayer model that 
the kink state is tightly localized around the center of the junction and that the associated $K$- and $K'$-valley currents are counterflowing with the same local distributions (apart from the overall sign),
thus leading to no orbital magnetization channel; the details of the analysis will be reported elsewhere.

\acknowledgments
 This work was supported in part by JSPS KAKENHI Grant No. 21K03534.

%\newpage

\appendix 

\section{Gauge transformation $G$}

In this appendix the gauge transformation $G$, introduced in Ref.~\cite{ks_edgeC_W}, 
is presented in a form suited for practical use. 
Here, for generality,  we follow the notation there, e.g., $v_{j}({\bf x}) \equiv e \ell a_{j}({\bf x})$ and 
$v\equiv e \ell (a_{y} + ia_{x})/\sqrt{2}$,
and write $v({\bf x})=e^{i \ell pZ^{\dag}} e^{i\ell p^{\dag}Z} \gamma_{\bf p} v({\bf r})$ with
center coordinate ${\bf r} = (i\ell^2\partial_{y_{0}}, y_{0})$ and  ${\bf p} = -i \nabla$; 
 $p= (p_{y} + ip_{x})/\sqrt{2}$, $ip =(\partial_{y} + i\partial_{x})/\sqrt{2}\equiv \partial$, 
 $\gamma_{\bf p}= e^{-{1\over{4}} \ell^2 {\bf p}^2}$, etc.
 (Potentials $a_{j}$ and $A_{0}$ may depend on time $t$, though left implicit.)
 We set $u({\bf r}) \equiv  \gamma_{\bf p} v({\bf r})$ and  $\ell \rightarrow 1$ below.

Let us first expand $v({\bf x})= e^{i pZ^{\dag}} e^{ip^{\dag}Z} u({\bf r})$ 
in powers of $(Z^{\dag}, Z)$ and write
\begin{equation}
v({\bf x})
= \Big[ F_{\xi} + \sum_{r=1}^{\infty}F^{r0}_{\xi} \partial^{r} 
+ \sum_{r=1}^{\infty}F^{0r}_{\xi}(\partial^{\dag} )^{r}\Big]  u({\bf r}), 
\end{equation}
where matrix operators 
\begin{equation}
F^{r0}_{\xi} = \sum_{s=0}^{\infty}{ (-\xi)^{s} \over{(s+r)! s!}}\, (Z^{\dag})^{s+r}Z^{s},\ \ 
F^{0r}_{\xi} = (F^{r0}_{\xi})^{\dag},
\end{equation}
and $F_{\xi} \equiv F^{00}_{\xi}$
are functions of derivative $\xi \equiv{1\over{2}}{\bf p}^2 = -\partial^{\dag}\partial$,
 \begin{equation}
(F^{r0}_{\xi})^{MN}  =  (F^{0r}_{\xi})^{NM}=\delta^{M, N+ r}\, \alpha_{N}^{r} \,  L_{N}^{r}(\xi),
\end{equation}
with $\alpha_{N}^{r} =  \sqrt{N!/(N+r)!}$;  $(F_{\xi})^{MN} =\delta^{MN} L_{N}(\xi)$.

 The gauge transformation $G = e^{iS}$  with $S \sim O(v)$
is cast in the form~\cite{ks_edgeC_W}
\begin{eqnarray}
S &=& -  \{(F_{\xi} -1)/2 \xi\}\,  (\nabla\cdot {\bf u}) 
\nonumber\\
&&+ F^{r0}_{\xi}\partial^{r-1} u({\bf r})
+  F^{0r}_{\xi}(\partial^{\dag})^{r-1} u^{\dag}({\bf r}),
\end{eqnarray}
where $\nabla\cdot {\bf u} \equiv {1\over{2}} \{\partial u^{\dag}({\bf r}) + \partial^{\dag}u({\bf r}) \}$;
the sum $\sum_{r=1}^{\infty}$ over $r$ is tacitly understood.  
Noting the basic commutators
\begin{eqnarray}
{[Z, F^{r,0}_{\xi}]} &\stackrel{r\ge 1}{=}&  F^{r-1,0}_{\xi},\  [Z, F_{\xi}] =  - \xi \, F^{01}_{\xi},
\nonumber\\
{[Z, F^{0r}_{\xi}]} &=& - \xi \, F^{0,r+1}_{\xi}, 
\end{eqnarray}
one can readily evaluate 
\begin{eqnarray}
[Z, S] &=& F_{\xi} u({\bf r}) +
 F^{01}_{\xi} (\nabla\cdot {\bf u})
 \nonumber\\
 &&+ F^{r 0}_{\xi}\partial^{r} u({\bf r})
+ F^{0,r+1}_{\xi} (\partial^{\dag})^{r} \partial u^{\dag}({\bf r}).
\end{eqnarray}

This leads to a transformed field  $v^{G} = v -[Z, S]$, to $O(v)$, 
of manifest electromagnetic gauge invariance,
\begin{equation}
v^{G} = 
-i \Big[ {1\over{2}} F^{01}_{\xi} \gamma_{\bf p} b_{z}({\bf r}) 
+ \sum_{r=2}^{\infty}F^{0r}_{\xi}(\partial^{\dag} )^{r-1}\gamma_{\bf p} b_{z} ({\bf r}) \Big],
\end{equation}
written with a magnetic field $b_{z}({\bf r}) \equiv -i (\partial v^{\dag} -\partial^{\dag}v)
= e \ell^2 \{ \partial_{x}a_{y}({\bf r}) -\partial_{y}a_{x}({\bf r})\}$ alone.
Note that $[v^{G}]^{NN}=0$ and 
\begin{equation}
i [v^{G}]^{N,N+r} = c_{N}^{r} L_{N}^{r}(\xi) (\partial^{\dag} )^{r-1}\gamma_{\bf p} b_{z} ({\bf r}),
\end{equation}
with $c_{N}^{1}= 1/(2\sqrt{N+1})$ and $c_{N}^{r} \stackrel{r\ge 2}{=} \sqrt{N! /(N+r)!}$.
Real fields $v_{x}^{G} = - \{ iv^{G} + (iv^{G})^{\dag} \}/\sqrt{2}$
and $v_{y}^{G}$ are naturally symmetric in orbitals $(M,N)$.

At the same time, $eA_{0}^{G} \equiv  G(eA_{0}({\bf x})  + \partial_{t})G$ is written as 
$eA_{0}^{G} = eA_{0} +\partial_{t}S + i[S, eA_{0} ] + O(u^2)$.
The lowest term $eA_{0}^{G;(u)} = eA_{0}({\bf x})  + \partial_{t}S$, to $O(u)$, 
is expressed in terms of $A_{0}$ and electric fields 
 $E_{j} = -\partial_{j}A_{0}({\bf r})  - \partial_{t} a_{j}({\bf r})$,
\begin{eqnarray}
A_{0}^{G;(u)} &=& \gamma_{\bf p} A_{0}({\bf r})  
-  \{(F_{\xi} -1)/2 \xi\}\,  \gamma_{\bf p}(\nabla\cdot {\bf E}) 
\nonumber\\
&& - \sum_{r=1}^{\infty}\Big [F^{r0}_{\xi} \partial^{r-1} \gamma_{\bf p} E 
+ F^{0r}_{\xi} (\partial^{\dag} )^{r-1}\gamma_{\bf p}E^{\dag} \Big],\ \  \ \ \ \ \ 
\end{eqnarray}
where $E= (E_{y} + iE_{x})/\sqrt{2}$.
For $A_{0}^{G;(uA)} \equiv  i[S, A_{0}] \sim O(u A_{0})$, we quote only the diagonal 
portion, to $O(dA_{0})$,
\begin{eqnarray}
A_{0}^{G;(uA)} &\stackrel{\rm diag}{=}& 
 i\Big[  (F_{\xi} u) \gamma_{\bf p}E^{\dag} -(F_{\xi} u^{\dag})\gamma_{\bf p}E \Big]
\nonumber\\
&=& - \{F_{\xi} {\bf u}({\bf r})\} \times\gamma_{\bf p} {\bf E}.
\end{eqnarray}

For (static) potentials $a_{x}(y)$ and $A_{0}(y)$ 
(that depend only on one coordinate $y$)
adopted in the main text,
the construction of $v_{x}^{G}$ and $A_{0}^{G}$ considerably simplifies.
Setting $u({\bf r}) \rightarrow  i u_{x}(y_{0})/\sqrt{2}$, 
${\bf p} \rightarrow p_{y}$ and $\partial \rightarrow d/\sqrt{2}$
(with $d \equiv \partial_{y_{0}}$) in $v^{G}$ above then leads to $[v_{x}^{G}]^{MN}$ 
quoted in Eq.~(\ref{vx_transG}).
The coordinate $y = y_{0} + \ell Y$ thereby takes a simple transformation law 
$G y G^{\dag} = y + \delta^{G} y + O(a_{x}^{2})$, 
with $\delta^{G} y = -i[Y,S] \sim O(a_{x})$ given by
\begin{equation}
\delta^{G} y = F_{\xi}\,  u_{x}(y_{0}) 
+  {1\over{2\sqrt{2}}} ( F^{10}_{\xi} +F^{01}_{\xi} ) d u_{x}(y_{0}). 
\end{equation}

As a result, $A_{0}(y)$ is transformed to 
\begin{equation}
A_{0}^{G}  \equiv GA_{0}(y)G^{\dag} = A_{0}(y + \delta^{G} y) + O(a_{x}^2).
\end{equation}
Note that $\delta^{G}y$ starts with the lowest multipole $u_{x}= \gamma_{p} v_{x}(y_{0})$. 
Let us isolate it and expand 
$A_{0}^{G}$ around $y \sim y_{0}^{u} \equiv y_{0} + u_{x}(y_{0})$ up to $O(a_{x} d^2 A_{0})$,
\begin{eqnarray}
A_{0}^{G} &=& A_{0}(y+ u_{x}) + (\Delta^{G} y) dA_{0}(y_{0}^{u}) 
\nonumber\\
&& + {\textstyle {1\over{2}}} \ell \{Y, \Delta^{G} y\} d^{2}A_{0}(y_{0}^{u}), 
\end{eqnarray}
with $\Delta^{G} y = \delta^{G} y -u_{x}$; 
it is understood that $d y_{0}^{u} =1$, i.e., $d$ acts on $y_{0}$ and not on  $u_{x}(y_{0})$.
In components $[A_{0}^{G}]^{MN} = [A_{0}^{G}]^{NM}$ and to $O(d^2A_{0})$,
 \begin{eqnarray}
 [A_{0}^{G}]^{NN}
&=&\! L_{N}(\xi)A_{0}^{u} + dA_{0}^{u} \{L_{N}(\xi)-1\}u_{x}
 \nonumber\\
&&+ {1\over{4}}\,  d^{2}A_{0}^{u}\big\{ L_{N-1}^{1}(\xi) + L_{N}^{1}(\xi) \big\} du_{x}, 
\nonumber\\
{[A_{0}^{G}]}^{N+1,N} \!\!
&=& c^{1}_{N} dA_{0}^{u}\,  \Big\{1 + {1\over{2}}\, L_{N}^{1}(\xi) du_{x} \Big\}
\nonumber\\
&&  + \zeta_{N}\,d^{2}A_{0}^{u} \{ L_{N}(\xi) +L_{N+1}(\xi)-2\}u_{x},
\nonumber\\
{[A_{0}^{G}]}^{N+2, N}\!\!
&\propto& d^2 A_{0}^{u} \{1 + O(du_{x}) \},  
\label{Azero_G_N}
\end{eqnarray}
where $A_{0}^{u}$ stands for  $\gamma_{p}A_{0}(y_{0}^{u}) =  \gamma_{p}A_{0}(y_{0} + u_{x})$ 
and $du_{x} = -\gamma_{p}b_{z}(y_{0})$; 
$c^{1}_{N} =1/ \sqrt{2(N+1)}$ and  $\zeta_{N} =  {1\over{2}} (N+1) c^{1}_{N}$.
In Appendix B, we use these formulas to calculate 
multipoles $[\mu^{G}(y)]^{MN}$ of interlayer  bias 
$\mu^{G}(y) \equiv \mu (GyG^{\dag})$.

\section{Response of kink states}

In this appendix we examine the PZM Hamiltonian $H$ with 
${\cal H}^{G} = {\cal H}_{0} + {\cal V}^{G}_{v} +  {\cal V}^{G}_{\mu}$,
quoted with ${\cal V}^{G}_{\mu}$ in Eq.~(\ref{V_mu_G}),
and derive the $O(a_{x})$ response of the kink states. 
The basic formulas for $[\mu^{G}(y)]^{MN}= [\mu (GyG^{\dag})]^{MN}$ 
are presented in Eq.~(\ref{Azero_G_N}). 
As remarked there, the lowest pole $u_{x}=\gamma_{p}v_{x}(y_{0})$ 
comes with $y_{0}$. We thus expand $[\mu^{G}(y)]^{MN}$ in multipoles of $m_{0}(y_{0} + u_{x})$
up to $O(d^{2}m_{0})$, with full multipoles of $v_{x}(y)$ kept. 
In the PZM sectors $N = (0,1)$ they read 
\begin{eqnarray}
[\mu^{G}]^{00}\!\!  &=&  \bar{m}_{0}+ {1\over{4}}\, d^2 \bar{m}_{0}\,  d u_{x},
\nonumber\\
{[\mu^{G}]}^{11}\!\!  
 &=& L_{1}(\xi) \bar{m}_{0} +   d \bar{m}_{0} L_{1}^{-1}(\xi)u_{x} 
 + {1\over{4}}\,d^2 \bar{m}_{0}  L_{1}^{2}(\xi)d u_{x}, 
\nonumber\\
{[\mu^{G}]}^{10}\!\!
&=&\!\!  {1\over{\sqrt{2}}} \Big[d \bar{m}_{0}\Big\{1+ {1\over{2}}  du_{x}\Big\}
+ {1\over{2}}  d^2  \bar{m}_{0}L_{1}^{-1}(\xi)u_{x}  
\Big].\ \  
\end{eqnarray}
Here $\bar{m}_{0}$ stands for  $\bar{m}_{0} \equiv m_{0}(y_{0}^{u})$ 
with $y_{0}^{u} \equiv y_{0} + u_{x}(y_{0})$,  
$du_{x}= -  \gamma_{p}b_{z}(y_{0})$ and
$\xi =-{1\over{2}} d^2$;
$L_{1}^{\alpha}(\xi) = \alpha + L_{1}(\xi) = \alpha +1 - \xi$ and
$L_{1}^{-1}(\xi) = -\xi$.

Let us next extract the  $n = (0,1)$ components from 
$({\cal H}^{G})^{mn}=  ({\cal H}_{0} + {\cal V}^{G}_{v} +  {\cal V}^{G}_{\mu})^{mn}$.
Note that 
$[{\cal V}^{G}_{\mu}]^{00}/\omega_{c} = -[\mu^{G}]^{00}$ and 
$[{\cal V}^{G}_{\mu}]^{10}/\omega_{c} = - c_{1}[\mu^{G}]^{10}$ 
while $[{\cal V}^{G}_{\mu}]^{11}/\omega_{c}$ 
is given by $e_{1}$ in Eq.~(\ref{e_one_y})
with $\mu(y) \rightarrow \mu^{G}(y)$.
Similarly, 
$[{\cal V}^{G}_{v}]^{11}/\omega_{c} = -c_{1}b_{1}\gamma_{p}b_{z}(y_{0})$ 
with $b_{1}$ fixed in Eq.~(\ref{b_one_y}), 
and  $[{\cal V}^{G}_{v}]^{00}= [{\cal V}^{G}_{v}]^{10}=0$.
The result is
\begin{equation}
({\cal H}^{G})^{mn} /\omega_{c} = h^{mn}(y_{0}^{u}) + \delta h^{mn}.
\end{equation}
Here $h^{mn}(y_{0}^{u})$ stands for $h^{mn}$ in Eq.~(\ref{h_pzm}) 
with $y_{0} \rightarrow  y_{0}^{u}$,
and $\delta h^{mn}= \delta h^{nm}$ consists of  the $O(du_{x}, d^2 u_{x})$ response,
\begin{eqnarray}
\delta h^{00} &=& {c_{1} \over{2\sqrt{2}\, g^2}}\, dR\, du_{x},
\nonumber\\ 
\delta h^{11} 
&=&  (c_{1}/g)^2 \Big[ \sqrt{2}\,c_{1}  R\, L_{1}^{-1}(\xi)u_{x}    + (h_{b})^{11} du_{x}\Big],
\nonumber\\
(h_{b})^{11}  &=&
 - 2 (c_{1})^2 Q  +  {c_{1}\over{2\sqrt{2}}}\, dR\, \Big\{L_{1}^{4}(\xi) -{1\over{g^2}}\Big\},
\nonumber\\
\delta h^{10} &=&  
{1\over{2}} (c_{1}/g)^2 \Big[ R\,  du_{x} + dR\, L_{1}^{-1}(\xi)u_{x} \Big],
\end{eqnarray}
where $Q=Q(y_{0})$,  $R =R(y_{0})$, etc.
As before, $h^{mn}(y_{0}^{u})$ is diagonalized by use of the unitary matrix $U$ in Eq.~(\ref{Rotation_U}), 
now with $U=U(y_{0}) \rightarrow U(y_{0}^{u})$, 
thus leading to the spectra $\hat{e}_{n}(y_{0}+ u_{x})$.
The $O(u_{x})$ response of level $n=(0,1)$ is thereby read from $(U\delta h U^{\dag})^{nn}$. 
The resulting spectra and response are summarized as 
$\hat{e}_{n}^{G}[y_{0}]$ in Eq.~(\ref{hat_eG_n}).

%%%%%%%%%%%%%%%%%%% References  trim %%%%%%%%%%%%%%

\end{document}